\documentclass[journal]{IEEEtran}
\usepackage{graphicx}
\usepackage{amsmath,amssymb,}

\ifCLASSINFOpdf
 \else
\fi
\begin{document}
%
\title{Negative $\beta$-encoder}
%
%
%

\author{Tohru~Kohda,~\IEEEmembership{Member,~IEEE,}
Satoshi~Hironaka, 
        and~Kazuyuki~Aihara
\thanks{ T. Kohda and  S. Hironaka are with the Department
of Computer Science and Communication Engineering, Kyushu University
Motooka 744, Nishi-ku, Fukuoka-city 819-0395, Japan, 
 Phone +81-92-802-3623, Fax +81-92-802-3627, (email: kohda@csce.kyushu-u.ac.jp)}
\thanks{K. Aihara is with the Department of Mathematical Engineering and
Information Physics, Faculty of Engineering, Tokyo University
and Aihara Complexity Modeling Project, ERATO, JST}
}

%
%

\markboth{Journal of \LaTeX\ Class Files,~Vol.~6, No.~1, January~2007}%
{Shell \MakeLowercase{\textit{{\it et al.}}}: Bare Demo of IEEEtran.cls for Journals}
%



\maketitle

\begin{abstract}
A new class of
 analog-to-digital (A/D) and digital-to-analog (D/A) converters 
using a flaky quantiser, 
called the $\beta$-encoder, has been shown to have exponential bit rate 
accuracy  while possessing a self-correction property for 
fluctuations of the amplifier factor $\beta$ and the quantiser threshold $\nu$.
 The probabilistic behavior of  such a flaky quantiser 
is  explained as  the deterministic dynamics of the multi-valued R\'enyi map. 
That is, a sample $x$ is always confined to a contracted 
subinterval while successive approximations of $x$ are   performed 
using  $\beta$-expansion even if $\nu$ may vary at each iteration. 
This viewpoint 
enables us to get the decoded sample, which is  equal to the midpoint of 
the subinterval, and its associated characteristic equation 
for recovering $\beta$  
which improves the quantisation error by more than 
$3\mbox{dB}$ when $\beta>1.5$.
The invariant subinterval under the R\'enyi map shows that 
$\nu$ should be set to around the midpoint of its associated greedy and 
lazy values. 
Furthermore, a new A/D converter is introduced called the negative $\beta$-encoder,  which further improves the quantisation error of the $\beta$-encoder.
A two-state Markov chain describing the $\beta$-encoder 
suggests
that a negative eigenvalue of its associated transition probability matrix 
reduces the quantisation error.
\end{abstract}


%
\IEEEpeerreviewmaketitle


\section{Introduction}
%
%
%
%

\IEEEPARstart{S}{AMPLING} and quantisation are necessary in almost all 
signal processing. 
The combined operations are called {\it analog-to-digital} (A/D) 
{\it conversion}. 
Since A/D converters are analog circuits, 
they have the fundamental problem that
instability of the circuit elements  degrades the A/D conversion.
There are a number of possible remedies to cope with this problem.

The standard sampling theorem states that if 
a band-limited signal $f(t)$ is sampled  at rates far above the Nyquist rate, 
called {\it oversampling}, 
then  it can be reconstructed 
from its samples, denoted by  
$\{f(\dfrac{n}{M})\}_{n\in {\cal Z}}$ (with $M>1$), 
 by the use of the following formula 
with an appropriate function $\varphi(t)$,
~\cite{Gunturk2k1,Gunturk2k3,Cvetkovic}
\begin{equation}
f(t)=\dfrac{1}{M}\sum_{n\in {\cal Z}} f(\dfrac{n}{M})
\varphi(t-\dfrac{n}{M}).
\label{sampling theorem oversampling ver.}
\end{equation}
The above formula does not result in a loss of information.
However, since the amplitudes of the samples are continuous variables, 
each sample is quantised according to amplitude 
into a finite number of levels. 
This quantisation process necessarily introduces  some distortion 
into the output. The magnitude of this type of distortion 
depends on the method by which the quantisation is performed.

Various kinds of A/D and digital-to-analog (D/A) conversions 
 have been proposed. Related topics include one-bit coding through 
over-sampled data~\cite{InoseYasuda} and high-quality AD conversions using a coarse quantiser together with feedback~\cite{Candy}, the concept of \lq\lq {\it democracy}"~\cite{Calderbank}, in which the individual bits in a coarsely quantised  representation of a signal are all given \lq\lq equal weight" in the approximation to the original signal, a pipelined AD converter~\cite{LewisGray}, 
 and a single-bit oversampled AD conversion using irregularly spaced 
 samples~\cite{Cvetkovic}. 

Given a  bandlimited function $f$, the $L$-bit pulse code modulation 
(PCM)~\cite{PCM} simply uses each sample value $f(n/M)$ with $L$ bits: 
one bit for its sign, followed by the first $L-1$ bits of the binary 
expansion of $f(n/M)$. 
It is possible to show that for a bandlimited signal, this algorithm achieves  
precision of order $O(2^{-L})$.
%
On the other hand, $\Sigma \Delta$ modulation
~\cite{Gunturk2k1,Gunturk2k3,Gray87,Gray89,Daubechies2k3}, 
another commonly implemented quantisation algorithm for a bandlimited 
function, achieves  precision that decays like an inverse polynomial
in the bit budget $L$. For example, a $k$th-order $\Sigma \Delta$
scheme produces an approximation where the distortion is of the order $O(L^{-k})$.
Although PCM is superior to $\Sigma \Delta$ modulation in  its level 
of distortion for a given bit budget, $\Sigma \Delta$ modulation 
has  practical features for analog circuit implementation. 
One of the key features is a ceratain self-correction property for quantiser 
threshold errors (bias) that is not shared by PCM.
This is one of several reasons why $\Sigma \Delta$ modulation is preferred for A/D conversion in practice.

In 2002, Daubechies {\it et al.}~\cite{Daubechies2k2} introduced 
a new architecture for A/D converters called the \textit{$\beta$-encoder} 
and showed the interesting result that it has exponential accuracy 
even if the $\beta$-encoder is iterated at each step in successive 
approximation of each sample using an imprecise quantiser with 
a quantisation error  and an offset parameter. 
Furthermore, in a subsequent paper~\cite{Daubechies2k61}, 
they introduced a \lq\lq flaky" version of an imperfect quantiser, 
defined as 
\begin{equation}
Q^f_{[\nu_0,\nu_1]}(z)=
\left\{
\begin{array}{lcr}
0,&\mbox{if}\,z\leq \nu_0,\\
1,&\mbox{if}\,z\geq \nu_1,\\
0\,\mbox{or}\,1,&\mbox{if}\,z\in(\nu_0,\nu_1),
\end{array}
\right.\label{eq:flaky}
\end{equation}
where $1\leq\nu_0<\nu_1\leq(\beta-1)^{-1}$ 
and made the remarkable observation that \lq\lq greedy" and \lq\lq lazy" as well as \lq\lq cautious"\footnote{Intermediate expansions~\cite{Dajani} between the greedy and lazy expansions are called \lq\lq cautious" by Daubechies {\it et al.}~\cite{Daubechies2k61}.}  expansions in the $\beta$-encoder with such a flaky quantiser exhibit exponential accuracy in the bit rate.
This $\beta$-encoder was a milestone in oversampled A/D and D/A conversions 
in the sense that it may become a good alternative to PCM.  
The primary reason is that the $\beta$-encoder consists of an analog circuit, 
with an amplifier with the factor $\beta$, a single-bit quantiser 
with the threshold $\nu$  and a single feedback loop for 
successive $L$-bit quantisation of  each sample
which uses the bit-budget efficiently, like PCM.
The $\beta$-encoder further guarantees the robustness of both 
$\beta$ and $\nu$ against fluctuations like $\Sigma \Delta$ modulation.
Nevertheless, it provides a simple D/A conversion using the estimated $\beta$ 
without knowing the exact value of $\beta$ with its offset 
in the A/D conversion~\cite{Daubechies2k62}.

This paper is devoted to dynamical systems theory 
for studying  ergodic-theoretic and probabilistic  properties 
of the $\beta$-encoder 
as a nonlinear system with feedback.  
We emphasize here that the flaky quantiser $Q^f_{\Delta_\beta}(\cdot)$ is 
exactly realized by the multi-valued R\'enyi map ({\it i.e.,} 
$\beta$-transformation)~\cite{Renyi} on the middle interval 
$\Delta_\beta=[\beta^{-1},\beta^{-1}(\beta-1)^{-1}]$ 
 so that probabilistic behavior in the \lq\lq flaky region" $\Delta_\beta$ is 
completely explained using  dynamical systems theory. 
 Our purpose is to give a \lq\lq dynamical" version of Daubechies {\it et al.}'s proof for the exponential accuracy of the $\beta$-encoder as follows. 
We can observe that a sample $x$ is always  confined 
to a subinterval of the contracted interval 
defined in this paper while  the successive approximation of $x$ is {\it stably}\footnote{A small real-valued  quantity, approximately proportional to the 
quantisation error, does not necessarily converge to 
any fixed value, {\it e.g.,} $0$ but may oscillate without diverging as 
discussed later in detail. 
Such a phenomenon is sometimes referred to as 
{\it chaos}.} 
performed using $\beta$-expansion 
  even if $\nu$ may vary at each iteration.
This enables us to obtain  the decoded sample easily, as it is equal to the 
midpoint of the subinterval, and it also  yields the characteristic equation 
for recovering $\beta$  which improves the quantisation error 
by more than $3\mbox{dB}$ over Daubechies {\it et al.}'s bound 
when $\beta>1.5.$ 
Furthermore, two classic $\beta$-expansions, known as the {\it greedy} and 
{\it lazy} expansions are proven to be perfectly symmetrical in terms of their 
quantisation errors. 
The invariant subinterval of the R\'enyi map further suggests that $\nu$ should be set to around the midpoint of its associated greedy and lazy values.
This paper presents a radix expansion of a real number in a 
negative real base, called  {\it a negative $\beta$-expansion} 
and a {\it negative $\beta$-encoder} 
in order to make  stable analog circuit implementation easier.
 Finally we observe a clear difference 
between a sequence of independent and identically distributed (i.i.d.) 
binary random variables generated by PCM and  
a binary sequence generated by the $\beta$-encoder 
based on the viewpoint that if the latter sequence is regarded as a $2$-state 
Markov chain with a $2\times2$ transition probability matrix, 
 then the matrix has a negative eigenvalue.

First, we  survey PCM from the viewpoint of dynamical systems 
because it is a typical example of a nonlinear map. 
PCM is an A/D converter that realizes binary expansion in the analog world. 
The binary expansion of  a given real number $r \in [-1, 1]$ has the form 
\begin{equation}
r = b_{0,B} \sum^{\infty}_{i=1} b_{i, B} 2^{-i},
\end{equation}
where $b_{0,B} = b_{0, B}(r) \in \{-1, 1\}$ is the sign bit, and 
$b_{i, B} = b_{i, B}(r) \in \{0,1\}\ i \ge 1$ are the binary digits of $|r|$.
We define the quantiser function $Q_1(\cdot)$ as
\begin{equation}
Q_1(x) =  \left\{
\begin{array}{ll}
0,      & x < 1,\\
1,      & x \ge 1.
\end{array}
\right.
\end{equation}
Then we have $b_{i, B}$ which can be computed by the following algorithm.
Let $u_1 = 2|r|$; 
the first bit $b_{1,B}$ is then given by $b_{1,B} = Q_1(u_1)$.
 The remaining bits are determined recursively: 
if $u_i$ and $b_{i,B}$ have been given, then we can define 
\begin{equation}
u_{i+1} = 2 (u_i - b_{i,B})
\end{equation}
and
\begin{equation}
b_{i+1,B} = Q_1(u_{i+1}),
\end{equation}
respectively. Such a sequence is also obtained with the 
Bernoulli shift map $B( x )$~\cite{Billingsley,Lasota,Boyarsky}, 
defined by
\begin{equation}
B(x) = 2x\ \mbox{mod}\ 1 =  \left\{
\begin{array}{ll}
2x,      & x < 1/2,\\
2x - 1, & x \ge 1/2,\\
\end{array}
\right.
\label{eq:Bernoulli}
\end{equation}
and its associated bit sequence $b_{i,B}(i=1,2,\ldots)$, defined by
\begin{equation}
b_{i,B} =  \left\{
\begin{array}{ll}
0,      & B^{i-1}(x) <  1/2,\\
1,      & B^{i-1}(x)  \ge 1/2.\\
\end{array}
\right.
\end{equation}
Iterating $B(x)$ for $x \in [0, 1)$ gives 
\begin{equation}
B^{i+1}(x) = 2B^i(x) - b_{i,B},\, i = 1, 2, \cdots, L \in \mathbb{N}.
\end{equation}
Then
\begin{eqnarray}
B^L(x) &=& 2(\cdots2(2x-b_{1,B})-b_{2,B}\cdots)-b_{L,B} \nonumber\\
         &=& 2^L x - \sum^L_{i=1}  b_{i,B} 2^{L-i}
\label{B^L(x)}
\end{eqnarray}
or
\begin{eqnarray}
x = \sum^L_{i=1} b_{i,B} 2^{-i} + 2^{-L}B^L(x).
\label{binary expansion1}
\end{eqnarray}
Hence 
$2^{-L}B^L(x) = 0$ as $L \rightarrow \infty$ because $B^L(x) \in [0, 1)$. 
That is, we get the binary expansion of $x$:
\begin{eqnarray}
x = \sum^{\infty}_{i=1} b_{i, B} 2^{-i}.
\end{eqnarray}
Suppose that a threshold shift $\rho$ occurs. 
Let $B_{\rho}(x)$ be the resulting map: 
\begin{equation}
B_{\rho}(x) =   \left\{
\begin{array}{ll}
2x,      & x < (1 + \rho)/2, \\
2x - 1, & x \ge (1 + \rho)/2, \\
\end{array}
\right.
\end{equation}
and $b_{i,B_{\rho}}(i=1,2,\ldots)$  its bit sequence:
\begin{equation}
b_{i,B_{\rho}}=  \left\{
\begin{array}{ll}
0,      & B_{\rho}^{i-1}(x) < (1+ \rho)/2,\\
1,      & B_{\rho}^{i-1}(x) \ge (1+ \rho)/2.\\
\end{array}
\right.
\end{equation}
Then we have its associated binary expansion of x, defined as 
\begin{equation}
x = \sum^{L}_{i=1} b_{i,B_{\rho}} 2^{-i} + 2^{-L} B_{\rho}^L(x).
\end{equation}
When $\rho>0$, we have  
$B_{\rho}(x):[0,1) \rightarrow [0, 1+ \rho)$ for $x \in [0, 1)$. 
 Iterating  $B_{\rho}(x)$ $L$ times gives 
$B^L_{\rho}(x):[0,1) \rightarrow [0, 1+ 2^{L-1} \rho)$. 
Thus we have
\begin{equation}
0 \le x - \sum^{L}_{i=1} b_{i, B_{\rho}} 2^{-i} <  2^{-L} + \dfrac{\rho}{2}.
\label{rho+}
\end{equation}
Conversely, suppose that $\rho<0$; then 
we get $B_{\rho}(x):[0,1) \rightarrow [\rho, 1)$ for $x \in [0, 1)$. 
Iterating  $B_{\rho}(x)$ $L$ times gives 
$B^L_{\rho}(x):[0,1) \rightarrow [2^{L-1} \rho, 1)$ which implies that
\begin{equation}
\dfrac{\rho}{2} \le x - \sum^{L}_{i=1} b_{i, B_{\rho}} 2^{-i} \le  0.
\label{rho-}
\end{equation}
Both (\ref{rho+}) and (\ref{rho-}) show that an A/D conversion does not work 
well because the quantisation errors don't decay.
\begin{figure}[b]
\begin{center}
\includegraphics[scale=0.23]{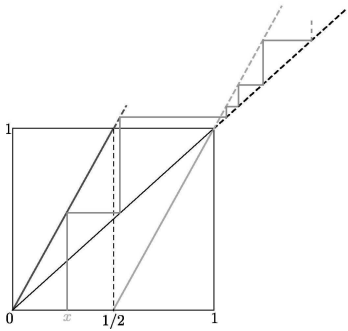} 
\end{center}
\caption{The divergence of a value $x$ in PCM when there is a threshold shift $\rho > 0$.}
\label{PCM_rho}
\end{figure}
Figure \ref{PCM_rho} shows the divergence of a value $x$ in PCM when there 
is a  threshold shift $\rho > 0$.
Such a map must be a mapping interval into interval (or at least a mapping 
interval onto interval)  so that the A/D conversion operates normally.  
 Fluctuations of the threshold are inevitable because 
every A/D converter is implemented as an analog circuit. 

However, $\beta$-encoders which realize \textit{$\beta$-expansion} 
using the expansion  by $\beta \in (1,2)$ as a radix, 
overcome this problem.
\begin{figure}[t]
\begin{center}
\includegraphics[scale=0.3]{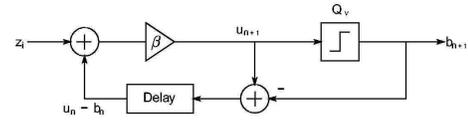} 
\end{center}
\caption{A $\beta$-encoder: for input $z_0 = y \in [0, 1), z_i = 0, i >1$, and 
\lq\lq initial conditions" $u_0 =
0$ and $b_{0, \beta} = 0$, the output $(b_{i, \beta})_{i \ge 1}$ gives the $\beta$-expansion for $y$
defined by the quantiser $Q_{\nu} = Q_1(\cdot/\nu)$,
with $\nu \in [1, (\beta -1)^{-1}]$.
This provides the \lq\lq greedy" and \lq\lq lazy" schemes for $\nu = 1$ and 
$\nu = (\beta - 1)^{-1}$, respectively. The $\beta$-encoder with 
$\beta=2$ and $\nu=1$ gives PCM.}
\label{BetaEncoder}
\end{figure}
%
The block diagram of the $\beta$-encoder is shown in 
Fig.\ref{BetaEncoder} with 
the  amplifier $\beta \in (1, 2)$ and the quantiser $Q_{\nu}$.
The quantiser $Q_{\nu}$ is defined by
\begin{equation}
Q_\nu(x) =  \left\{
\begin{array}{ll}
0,      & x < \nu,\\
1,      & x \ge \nu,
\end{array}
\right.\nu \in [1, (\beta -1)^{-1}].
\end{equation}
Note that  the $\beta$-encoder with $\beta=2$ and $\nu=1$  provides the PCM.
The bit sequences $b_{i,\beta}$ can be calculated recursively as follows.
Let $u_1 = \beta x$;
the first bit $b_{1,\beta}$ is then given by $b_{1,\beta}=Q_{\nu}(u_1)$.
The remaining bits are obtained recursively;
given $u_i$ and $b_{i,\beta}$, 
we define $u_{i+1}=\beta(u_i - b_{i,\beta})$ 
and $b_{i+1,\beta} = Q_\nu(u_{i+1})$.
Daubechies  {\it et al.}~\cite{
Daubechies2k2,Daubechies2k61,Daubechies2k62} introduced 
the flaky quantiser, defined by Eq.(\ref{eq:flaky}) and gave  the 
important result that  the $\beta$-encoder can perform normally 
and has  exponential accuracy even if the quantiser threshold $\nu$ 
fluctuates over the interval $[1, (\beta - 1)^{-1}]$. 
\section{Basics of dynamical systems theory}
We deal exclusively with the {\it asymptotic behavior} exhibited 
by a dynamical system. 
In particular, we limit ourselves to a map of an interval $I$, 
called an {\it interval map}~\cite{Lasota,Boyarsky}. 
The  following short review of the fundamentals~\cite{BrucksBruin} 
is provided to explain the dynamics of the $\beta$-maps. 

Given $E\subset\mathbb{R}$ and the continuous map $\tau:E\rightarrow E$, 
the set $E$ and the map $\tau$ form a {\it dynamical system}, denoted by 
$(E,\tau)$. 
For a given $x\in I$ a sequence of forward iterates  
\begin{equation}
x,\tau(x), \tau^2(x)=\tau(\tau(x)),\tau^3(x)=\tau(\tau(\tau(x))),\ldots
\label{eq:systemeq}
\end{equation}
is referred to as the {\it forward trajectory} (or {\it  orbit}) of $x$. 
We call $F\subset E$ {\it invariant} if $\tau(F)\subset F$. 
We call $f:E_1\rightarrow E_2$ a {\it homeomorphism} if $f$ is one-to-one and both $f$ and $f^{-1}$ are continuous. If, in addition, $f$ is onto, we call $f$ an {\it onto homeomorphism}. 
Let $f:A\rightarrow A$ and $g:B\rightarrow B$ be given. We say that $f$ and $g$ are {\it topologically conjugate} if there is an {onto homeomorphism} 
$h:A\rightarrow B$ such that $h\circ f=g\circ h$.
The homeomorphism $h$ is called the {\it conjugacy} between $f$ and $g$. 
We say that a map $\tau$ with its invariant subinterval $J$ is 
{\it locally eventually onto} if for every $\varepsilon>0$ there exists 
$M\in\mathbb{N}$ 
such that, if $U$ is an interval with $|U|>\varepsilon$ and if $n\geq M$, 
then $\tau^n(U)=J$.

One of the main problems in dynamical systems theory is to 
describe the distribution of orbits. 
That is, we wish to know how the iterates of points under an interval map vary 
over the interval. Ergodic theory provides answers to such questions, 
particularly the notions of the ergodicity and the invariant measure. 
Let $\mu$ be an absolutely continuous invariant measure for the map $\tau$,
then we have the following theorem.\\
{\bf Birchoff Ergodic theorem}:~\cite{Lasota,Boyarsky,BrucksBruin}\\
 (i) 
Let $\tau$ be a {\it measure preserving} map of an interval $I$.
Then for any integrable function $h(x)$, the {\it time average} 
$\displaystyle{\lim_{N\rightarrow\infty}\frac1N\sum_{i=0}^{N-1}h(\tau^i(x))}$
exists for almost all $x\in I$ with respect to  $\mu$.\\
(ii) If, in addition, $\tau$ is {\it ergodic}, 
then the time average 
is equal to the {\it space average} 
$\int hd\mu$ for almost every $x$ with respect to  $\mu$.

The first important result on the existence of an absolutely continuous invariant measure is now considered to be a 
{\it folklore theorem} which originated with the basic result due to 
R\'enyi~\cite{Renyi}. His key idea has been used in more general proofs 
by Adler and Flatto\cite{AdlerFlatto}.

{\bf Definition $1$}
{\it 
Let $I$ be an interval and $\{I_i\}$ be a finite partition of $I$ into 
subintervals. Let $\tau:I\rightarrow I$ satisfy the following conditions:
\begin{enumerate}
\item {\it piecewise smoothness}, i.e., $\tau|_{I_i}$ has a $C^2$-extension to 
the {\it closure} $\overline{I}_i$ of $I_i$.
\item {\it local invertibility}, i.e., $\tau|_{I_i}$ is strictly monotonous.
\item {\it Markov property}, i.e., $\tau(\overline{I}_i)=$\,union of several 
$\overline{I}_j$.
\item {\it Aperiodicity}, i.e., there exists an integer $p$ such that 
$\tau^p(\overline{I}_i)=\overline{I}$ for all $i$.
\end{enumerate}
If 1)-3) hold, then $\{I_i\}$ is called a {\it Markov partition} for $\tau$ (or $\tau$ is a {\it Markov map} for $\{I_i\}$). ~\footnote{The Bernoulli shift map is a typical example of a map satisfying 1)-4).}
\label{def:Markov}
}

Condition 4) is added to ensure that the following theorem holds.\\
{\bf Folklore theorem}: {\it Assume that 1)-4) 
hold and that $\tau$ is {\it eventually expansive}, i.e., for some iterate $\tau^n$, $|d\tau^n/dx|\geq\theta>1$ for all $x$. Then
 $\tau$ has a finite Lesbesgue-equivalent measure $m$ and furthermore 
$dm=\rho(x)dx$, where $\rho(x)$ is piecewise continuous and $D^{-1}<\rho<D$ for some $D$.} 

Under conditions 1)-4), the converse of the folklore theorem also holds.
\section{Multi-valued R\'enyi map and flaky quantiser}
The $\beta$-expansion ($\beta > 1,\,\beta\not\in\mathbb{Z}$) is obtained 
as a basis of  the $\beta$-encoder  according to the classic ergodic 
theory~\cite{Renyi, Gelfond, Parry60, Parry64, Erdos,Erdos90}.
R\'enyi~\cite{Renyi} defined the $\beta$-transformation: $x\mapsto\beta x\mod1$ for real numbers $x\in(0,1]$ and  $\beta>1$.
Gelfond~\cite{Gelfond} and Parry~\cite{Parry60} gave its finite invariant measure. Parry~\cite{Parry64} defined the 
linear modulo one  transformation (or $(\beta,\alpha)$-transformation, 
a generalized R\'enyi map): $x\mapsto\beta x+\alpha\,\mod1$ 
for real numbers $x\in(0,1]$ and $\beta\geq1$, $0\leq\alpha<1$, and 
gave a finite invariant measure for a (strongly) ergodic linear modulo one 
transformation 
as follows.  \\
{\bf Parry's result}~\cite{Parry60,Parry64}:\,{\it If $\tau$ is a linear modulo one transformation ($\tau(x)=\beta x+\alpha \mod1,\,\beta>1,\,0\leq\alpha <1$), then $\nu(E)=\displaystyle{\int_E \rho(x)dx}$ is a finite signed measure invariant under $\tau$, where $\rho(x)$ is an unnormalized density given as 
\begin{eqnarray}
\rho(x)\hspace*{-3mm}&=&\hspace*{-4mm}\displaystyle{\sum_{x<\tau^n(1)}\gamma^n},\,\mbox{for}\,\alpha=0\, (~\cite{Gelfond,Parry60}),\label{eq.Parry1}\\
\rho(x)\hspace*{-3mm}&=&\hspace*{-4mm}\displaystyle{\sum_{x<\tau^n(1)}\gamma^n-\sum_{x<\tau^n(0)}\gamma^n},\,\mbox{for}\,\alpha\neq0 \,(~\cite{Parry64}), 
\end{eqnarray}
 with $\gamma=\beta^{-1}$. 
If $\tau$ is strongly ergodic, then $\rho(x)\geq0$ for almost all $x$ and $\nu$ is a finite positive measure invariant under $\tau$}.

\newcommand{\lorder}{\stackrel{\mathrm{L}}{<}}
Erd\"{o}s {\it et al.}~\cite{Erdos,Erdos90} showed that the $\beta$-expansion 
has multiple representations of a real number
$x \in [0, (\beta -1)^{-1})$ as follows. 
 They introduced the lexicographic order $\lorder$ on the real 
sequences:$(b_i) \lorder (b'_i)$ if there is a positive integer $m$ 
such that $b_i = b'_i$ for all $i<m$ and $b_m < b'_m$.
It is easy to verify that for every fixed $x$ with $0\leq x\leq(\beta-1)^{-1}$ 
in the set of all expansions of $x$,
there exist a maximum and a minimum with respect to this 
order, namely the so-called \textit{greedy} and \textit{lazy} expansions.
The greedy expansions were studied originally by R\'enyi~\cite{Renyi}, 
where they were called $\beta$-expansions. 
A number $x$ has a unique expansion if and only if its greedy and lazy 
expansions coincide.
Erd\"{o}s {\it et al.} defined the bit sequence of these expansions 
recursively as follows:
if $m \ge 1$ and if the bit sequence $b_i$ of the greedy expansion of $x$
is defined for all $i<m$, then we set
\begin{equation}
b_m = \left\{
\begin{array}{ll}
1,\ &  \sum_{i<m} b_i \gamma^{i} + \gamma^{m} \le x,\\
0, & \sum_{i<m} b_i \gamma^{i} + \gamma^{m} > x.
\end{array}\right.
\label{greedy}
\end{equation}
If $m \ge 1$ and if the bit sequence $b_i$ of the lazy expansion of $x$
is defined for all $i<m$, then we set
\begin{equation}
b_m = \left\{
\begin{array}{ll}
0,\ &   \sum_{i<m} b_i \gamma^{i} + \sum_{i>m}\gamma^{i} \ge x,\\
1,\ &   \sum_{i<m} b_i \gamma^{i} + \sum_{i>m}\gamma^{i} < x.
\end{array}\right.
\label{lazy}
\end{equation}
Erd\"{o}s {\it et al.}~\cite{Erdos90} noted the following duality of 
the greedy and lazy expansions.
Given $x\in[0, (\beta - 1)^{-1})$, we define  $\psi(x)$ 
by
\begin{equation}
\psi(x) = (\beta - 1)^{-1} - x.
\end{equation}
Using the trivial relation 
$\sum_{i<m} b_i \gamma^{i} + \sum_{i>m}\gamma^{i} 
=(\beta - 1)^{-1}-\sum_{i<m} \overline{b}_i\gamma^{i}-\gamma^{m},$
we can rewrite Eq.(\ref{lazy}) as
\begin{equation}
\hspace*{-12mm}b_m = \left\{
\begin{array}{ll}
0,\ &  (\beta - 1)^{-1} - \sum_{i<m} \overline{b}_i \gamma^{i} - \gamma^{m} \ge x,\\
1,\ &  (\beta - 1)^{-1} - \sum_{i<m} \overline{b}_i \gamma^{i} - \gamma^{m} < x, \end{array}\right.
\label{lazy2}
\end{equation}
or
\begin{equation}
\hspace*{-4mm} b_m = \left\{
\hspace*{-2mm} \begin{array}{ll}
0, &  \sum_{i<m} \overline{b_i} \gamma^{i}+\gamma^{m} 
\le
\psi(x),\\
1, &   \sum_{i<m} \overline{b_i} \gamma^{i} +\gamma^{m} 
>
\psi(x),
\end{array}
\right. 
\end{equation}
where $\overline{b}_i = 1 - b_i$. 
Introducing $c_i=\overline{b_i}$, we get the greedy expansion of $\psi(x)$:
\begin{equation}
\hspace*{-7mm} c_m = \left\{
\begin{array}{ll}
1,\ &  \sum_{i<m} c_i\gamma^{i}+\gamma^{m} 
\le\psi(x),\\
0,\ &   \sum_{i<m} c_i\gamma^{i}+\gamma^{m} 
>\psi(x),
\end{array}
\right. 
\end{equation}
which  has the dual roles of 
the greedy expansion $\{b_i\}$ of $x$ and the
lazy expansion $\{b_i\}$ of $x$, i.e., the greedy expansion 
$\{c_i\}$ of $\psi(x)$.

We now define several different kinds of dynamical systems 
governed by multi-valued R\'enyi maps 
on the middle interval $\Delta_\beta=[\gamma,\gamma(\beta-1)^{-1}]$ 
that realize 
 Daubechies {\it et al.}'s flaky quantiser $Q^f_{\Delta_\beta}(\cdot)$, 
defined by Eq.(\ref{eq:flaky}) as follows. 

\begin{figure}[htbp]
\begin{center}
\includegraphics[scale=0.45]{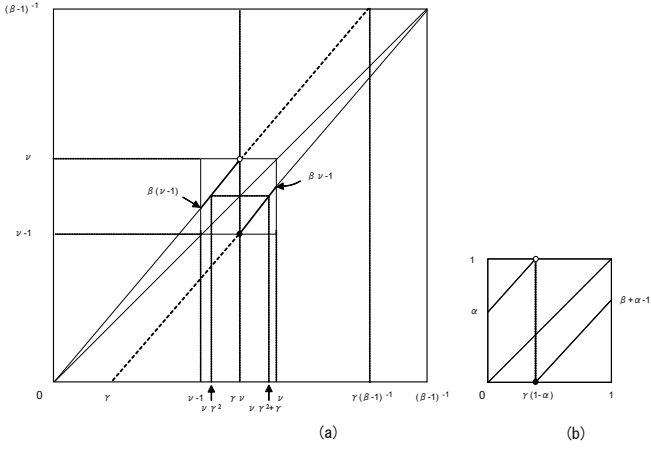} 
\caption{(a) \lq\lq {\it cautious}"-expansion map~\cite{Daubechies2k61}: 
$C_{\beta,\nu}(x)$ for $1<\nu<(\beta-1)^{-1}$, which is  
locally eventually onto $[\nu-1,\nu]$. 
Renormalizing the interval $[\nu-1,\nu]$ into the unit interval $[0,1]$ shows 
that such a locally eventually onto map is equivalent to 
the Parry's linear modulo one transformation (or $(\beta,\alpha)$-map) 
as shown in (b).}
\label{cautiousmap}
\end{center}
\end{figure}

Let  $C_{\beta, \nu}(x)$ be the cautious map, 
 shown in Fig.\ref{cautiousmap}(a), defined by
\begin{equation}
\hspace*{-6mm}
C_{\beta, \nu}(x) =  \left\{
\begin{array}{ll}
\beta x,      & x <\gamma\nu, \\
\beta x - 1, & x \ge \gamma\nu, \\
\end{array}
\right.\nu \in (1, (\beta -1)^{-1}),
\label{eq:Renyimap}
\end{equation}
which determines the flaky quantiser 
$Q^f_{[\gamma,\gamma(\beta-1)^{-1}]}(\cdot)$ 
 and gives its associated bit sequence 
$\{b_{i,C^i_{\beta,\nu}}\}_{i=1}^\infty$, 
defined by 
\begin{equation}
b_{i,C^i_{\beta,\nu}} =  \left\{
\begin{array}{ll}
0,      & C_{\beta, \nu}^{i-1}(x) < \gamma \nu, \\
1,      & C_{\beta, \nu}^{i-1}(x)  \ge \gamma\nu. \\
\end{array}
\right.
\label{eq:Renyimapbit}
\end{equation}
 Then we get the following cautious expansion of $x$ by the map 
$C_{\beta, \nu}(x)$ 
\begin{eqnarray}
x = \sum^{m-1}_{i=1} b_{i,C^i_{\beta,\nu}} \gamma^{i} + 
\gamma^{m-1}C_{\beta, \nu}^{m-1}(x),
\label{eq:cautious expansion}
\end{eqnarray}
which implies that each 
$x \in [0, (\beta-1)^{-1}]$ has a representation
\begin{equation}
x = \sum^{\infty}_{i=1} b_{i,C^i_{\beta,\nu}} \gamma^i,\quad b_{i,C^i_{\beta,\nu}}\in\{0,1\} 
\end{equation}
because $\gamma^{m-1}C_{\beta, \nu}^{m-1}(x) = 0$ when $m\rightarrow \infty$.
The cautious expansion map $C_{\beta,\nu}(x),\,\nu\in(1,(\beta-1)^{-1})$ 
with a unique  point of discontinuity $c=\gamma\nu$ 
has its strongly invariant subinterval $[\nu-1,\nu]$  because 
 the map $C_{\beta,\nu}(x)$ is {\it locally eventually onto}  
as shown in Fig.\ref{cautiousmap}(a).
This map defines its dynamical system, defined as 
$([\nu-1,\nu),C_{\beta,\nu}(x))$, 
which is illustrated by the bold lines in Fig.\ref{cautiousmap}(a). 

Let $\nu_1^i=\nu_1\cdots\nu_i$ 
be an abbreviation for a sequence of $i$ thresholds $\nu_j\in[1,(\beta-1)^{-1}],\,1\leq j\leq i$. Let $C_{\beta,\nu_1^i}^i$ and $b_{i,C_{\beta,\nu_1^i}^i}$ be 
the $i$-iterated map, recursively defined as 
\begin{equation}
\hspace*{-6mm}
C_{\beta,\nu_1^i}^i(x)=C_{\beta,\nu_i}(C_{\beta,\nu_1^{i-1}}^{i-1}(x))
=C_{\beta,\nu_i}(C_{\beta,\nu_{i-1}}(C_{\beta,\nu_1^{i-2}}^{i-2}(x)))
\end{equation}
and its associated binary random variable, defined as
\begin{equation}
b_{i,C_{\beta,\nu_1^i}^i}=
\left\{
\begin{array}{lcr}
0,& C_{\beta,\nu_1^{i-1}}^{i-1}(x)<\gamma\nu_i,\\
1,& C_{\beta,\nu_1^{i-1}}^{i-1}(x)\geq\gamma\nu_i,
\end{array}
\right.
\end{equation}
respectively.
Then the cautious expansion of $x$ by 
the map $C_{\beta, \nu}(x)$ using the threshold sequence $\nu_1^L$ is 
\begin{eqnarray}
x = \sum^{L}_{i=1} b_{i,C^i_{\beta,\nu_1^i}} \gamma^{i} + 
\gamma^{L}C_{\beta, \nu_1^L}^{L}(x),
\label{eq:cautious expansion1}
\end{eqnarray}
{\it i.e.,} the onto-mapping relation $0\leq C^L_{\beta,\nu_1^L}(x)\leq(\beta-1)^{-1}$ tells us that the sample $x$ is always confined to the $L$th stage 
subinterval, defined by 
\begin{equation}
I_{L,C^L_{\beta,\nu_1^L}}(x)  = [
\sum^{L}_{i=1}b_{i,C^i_{\beta,\nu_1^L}} \gamma^i, 
\sum^{L}_{i=1}b_{i,C^i_{\beta,\nu_1^L}} \gamma^i+(\beta-1)^{-1} \gamma^L), 
\label{eq:varying cautious expansion interval}
\end{equation}
where successive approximations to $x$ in the $\beta$-expansion  are performed 
using $\nu_1^L$. 
This elementary observation is important in discussing 
the contraction process of the interval by $\beta$-expansion. 
In order to avoid shortcuts in proving the above observation, 
it is worthwhile to discuss the quantiser in two different situations 
separately: the case  where a fixed sequence $\nu_1^L=\underbrace{\nu^*\cdots\nu^*}_{L\,\mbox{times}},\,\nu^*\in[1,(\beta-1)^{-1}]$ is used and 
the one where a varying sequence $\nu_1^L=\nu_1\cdots\nu_i\cdots\nu_L,\,\nu_i=\nu^*(1+u_i)\in[1,(\beta-1)^{-1}],\,1\leq i\leq L$ with a bounded random fluctuation $u_i$ is used.

Assume for simplicity that the quantiser threshold 
$\nu^*\in[1,(\beta-1)^{-1}]$ is fixed. 
Let us consider the special case where 
$\nu_i=1$ (or $\nu_i=(\beta-1)^{-1}$), $\,1\leq i\leq L$  
then it is called the \lq\lq{\it steady}" greedy (or lazy) expansion, 
{\it i.e.,} the classic greedy (or lazy) expansion. 
 
Let  $C_{\beta,1}(x)$ be the greedy expansion map, defined by
\begin{equation}
C_{\beta,1}(x) = 
\left\{
\begin{array}{ll}
\beta x,       & x < \gamma, \\
\beta x - 1, & x \ge \gamma, \\
\end{array}
\right.
\end{equation}
and  $b_{i,C^i_{\beta,1}}$ its associated bit sequence,  
defined by 
\begin{equation}
b_{i,C^i_{\beta,1}} = 
\left\{
\begin{array}{ll}
0,       & C_{\beta,1}^{i-1}(x) < \gamma, \\
1, & C_{\beta,1}^{i-1}(x) \ge \gamma, 
\end{array}
\right. i=1,2,\ldots.
\label{bT(x)}
\end{equation}
Then the following is the greedy expansion of $x$ by the map $C_{\beta,1}(x)$:
\begin{equation}
x = \sum^{m-1}_{i=1} b_{i,C^i_{\beta,1}} \gamma^i + 
\gamma^{m-1} C_{\beta,1}^{m-1}(x).
\label{eq:greedy expansion}
\end{equation}
Equivalently, we have 
\begin{equation}
\sum_{i<m} b_{i,C^i_{\beta,1}}\gamma^i=x-\gamma^{m-1} C_{\beta,1}^{m-1}(x) 
\label{eq:greedy}
\end{equation}
 which enables us to rewrite Eq.(\ref{bT(x)}) as follows:
\begin{eqnarray*}
&&
b_{m,C^m_{\beta,1}} = \left\{
\begin{array}{ll}
1,\ &  C_{\beta,1}^{m-1}(x) \ge \gamma,\\
0,\ &  C_{\beta,1}^{m-1}(x)  < \gamma,
\end{array}\right. \nonumber \\
\Leftrightarrow  &&
b_{m,C^m_{\beta,1}} = \left\{
\begin{array}{ll}
1,\ &  x -  \gamma^{m-1} C_{\beta,1}^{m-1}(x) + \gamma^{m} \le x,\\
0,\ &  x -  \gamma^{m-1} C_{\beta,1}^{m-1}(x)  + \gamma^{m} > x,
\end{array}\right. 
\nonumber \\
\Leftrightarrow  &&
b_{m,C^m_{\beta,1}} = \left\{
\begin{array}{ll}
1,\ &  \sum_{i<m} b_{i,C^i_{\beta,1}} \gamma^{i} + \gamma^{m} \le x,\\
0,\ &  \sum_{i<m} b_{i,C^i_{\beta,1}} \gamma^{i}  + \gamma^{m} > x.
\end{array}\right. 
\end{eqnarray*}
This suggests that $b_{m,C^m_{\beta,1}}$ 
in Eq.(\ref{bT(x)}) is equal to $b_m$ in Eq.(\ref{greedy}).

Let $C_{\beta,(\beta -1)^{-1}}(x)$ be  the lazy expansion map 
defined by
\begin{equation}
C_{\beta,(\beta -1)^{-1}}(x) = 
\left\{
\begin{array}{ll}
\beta x,       & x \le \gamma(\beta -1)^{-1}, \\
\beta x - 1, & x > \gamma(\beta - 1)^{-1}, \\
\end{array}
\right.
\end{equation}
and $b_{i,C^i_{\beta,(\beta -1)^{-1}}}(i=1,2,\ldots)$  its associated bit 
sequence, defined by
\begin{equation}
b_{i,C^i_{\beta,(\beta -1)^{-1}}} = 
\left\{
\begin{array}{ll}
0,       & C_{\beta,(\beta -1)^{-1}}^{i-1}(x) \le \gamma (\beta-1)^{-1},\\
1, & C_{\beta,(\beta -1)^{-1}}^{i-1}(x) > \gamma (\beta-1)^{-1}.\\
\end{array}
\right.
\label{bS(x)}
\end{equation}
Then the following is the lazy expansion of $x$ 
by the map $C_{\beta,(\beta -1)^{-1}}(x)$:
\begin{equation}
x = \sum^{m-1}_{i=1} b_{i,C^i_{\beta,(\beta -1)^{-1}}}\gamma^i 
+ \gamma^{m-1} C_{\beta,(\beta -1)^{-1}}^{m-1}(x)
\label{eq:lazy expansion}
\end{equation}
Equivalently, we get 
\begin{equation}
\sum_{i<m} b_{i,C^i_{\beta,(\beta -1)^{-1}}} \gamma^i = 
x -  \gamma^{m-1} C_{\beta,(\beta -1)^{-1}}^{m-1}(x) 
\label{eq:Lazy}
\end{equation}
 which enables us to rewrite Eq.(\ref{bS(x)}) as follows:
\begin{eqnarray*}
\hspace*{-9mm}
&&
b_{m,C^m_{\beta,(\beta -1)^{-1}}}\hspace*{-1mm}=\hspace*{-1mm} 
\left\{\hspace*{-1mm}
\begin{array}{ll}
0,\hspace*{-1mm}&\hspace*{-2mm}
  C_{\beta,(\beta -1)^{-1}}^{m-1}(x) \le \gamma (\beta-1)^{-1},\\
1,\hspace*{-1mm}&\hspace*{-2mm}
  C_{\beta,(\beta -1)^{-1}}^{m-1}(x)  > \gamma (\beta-1)^{-1},
\end{array}\right. \nonumber \\
\hspace*{-2mm}\Leftrightarrow  \hspace*{-6mm}
&&\hspace*{-1mm}
b_{m,C^m_{\beta,(\beta -1)^{-1}}}\hspace*{-1mm}=\hspace*{-1mm}
\left\{\hspace*{-1mm}
\begin{array}{ll}
0,\hspace*{-1mm}&\hspace*{-2mm}
  \sum_{i<m} b_{i,C^i_{\beta,(\beta -1)^{-1}}} \gamma^{i} + 
\sum_{i>m}\gamma^{i} \ge x,\\
1,\hspace*{-1mm}&\hspace*{-2mm} 
 \sum_{i<m} b_{i,C^i_{\beta,(\beta -1)^{-1}}} \gamma^{i} + 
\sum_{i>m}\gamma^{i} < x,
\end{array}\right. 
\end{eqnarray*}
because  $\sum_{i>m}\gamma^{i}=(\beta-1)^{-1}\gamma^m$. 
This suggests that $b_{m,C^m_{\beta,(\beta -1)^{-1}}}$ in Eq.(\ref{bS(x)}) 
is equal to $b_m$ in Eq.(\ref{lazy}).
%
\begin{figure}[htbp]
\begin{center}
\includegraphics[scale=0.45]{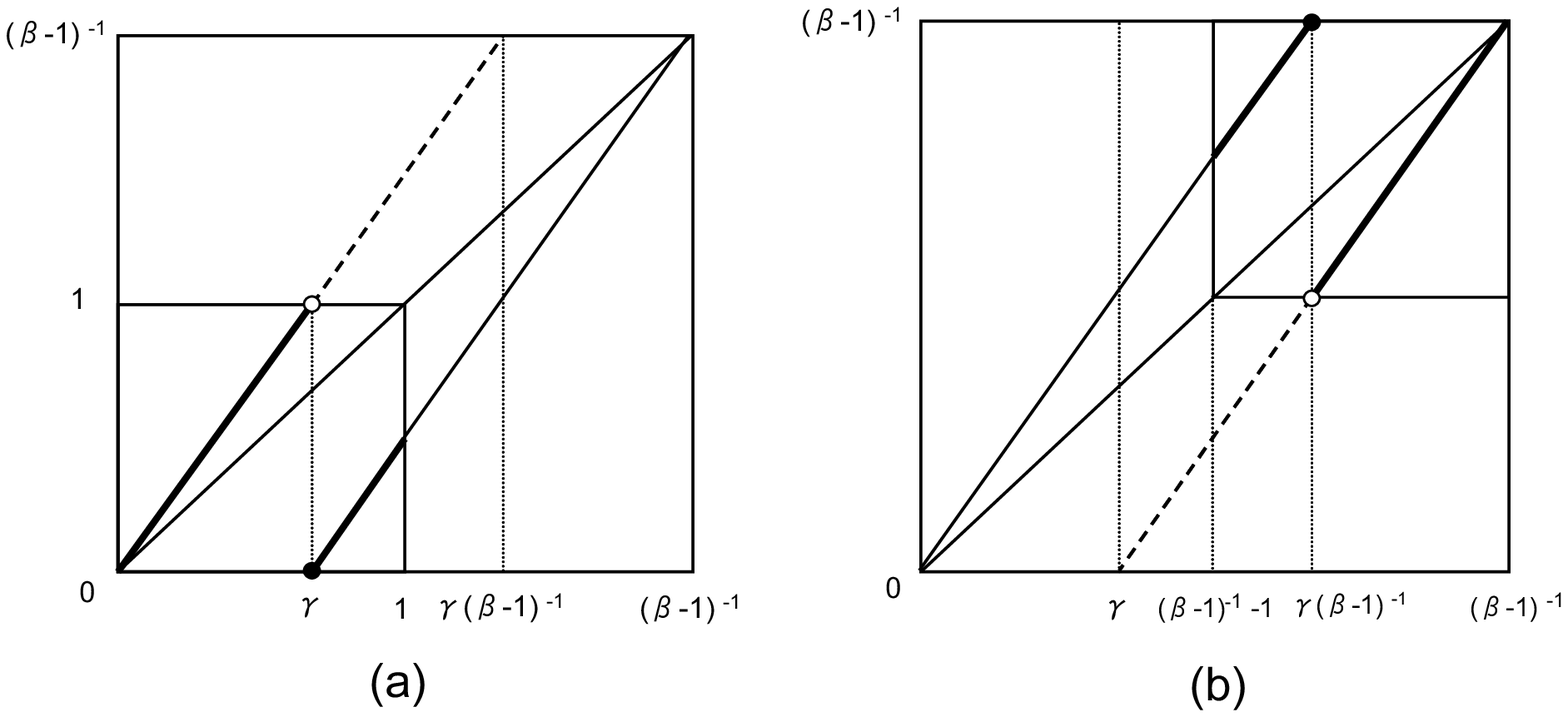} 
\caption{(a) Greedy-expansion map:$C_{\beta,1}(x)$, which is 
locally eventually onto $[0,1)$ and (b) lazy-expansion map:
$C_{\beta,(\beta-1)^{-1}}(x)$, which is locally eventually onto 
$[(\beta-1)^{-1}-1,\,(\beta-1)^{-1})$. These multivalued maps on the 
middle interval $\Delta_\beta=[\gamma,\gamma(\beta-1)^{-1}]$ corresponds 
exactly to the flaky quantiser $Q^f_{\Delta_\beta}(\cdot)$.}
\label{greedylazymap}
\end{center}
\end{figure}
%
\begin{figure}[htbp]
\begin{center}
\includegraphics[scale=0.45]{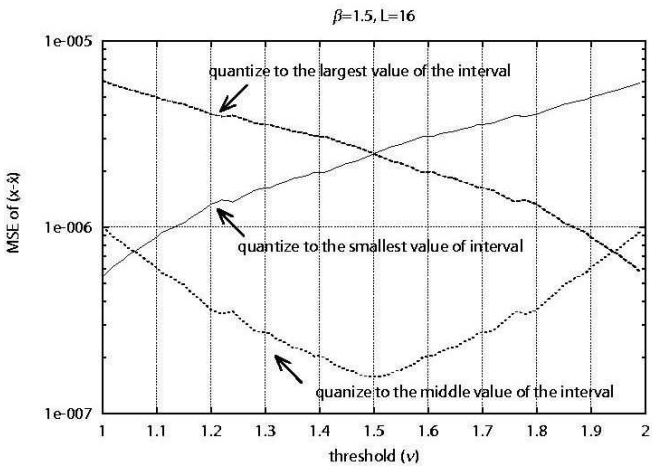} 
\end{center}
\caption{The $\mbox{MSE}
(\widehat{x}_{L,C^L_{\beta,\nu^*}}(\gamma,p_L))$,\,$0\leq p_L\leq2$ 
 using the exact $\beta$ of the $\beta$-encoder with  $\,\beta =1.5$, $L=16$, 
and fixed $\nu^*\in [1,(\beta-1)^{-1}]$.}
\label{Varxp_L=012}
\end{figure}
%

Let $\tau:I\rightarrow I$ be an interval map with a unique  point of discontinuity $c$ such that $\tau(c^-)=\lim_{t\uparrow c}\tau(t)$ is not equal to $\tau(c^+)=\lim_{t\downarrow c}\tau(t)$. 
To each $x\in I$ we associate an element of 
$\{0,1,\ast\}^\mathbb{N}$ by listing the sequence of adresses 
of the forward orbit of $x$, called 
 the {\it itinerary of $x$ under the map $\tau(x)$}, 
denoted as ${\bf s}_\tau(x)=(s_0,s_1,\cdots,s_n,\ldots)$, 
is defined by:
\begin{equation}
s_j=
\left\{
\begin{array}{lcr}
0,&&\mbox{if}\,\tau^j(x)<c,\\
\ast,&&\mbox{if}\,\tau^j(x)=c,\\
1,&&\mbox{if}\,\tau^j(x)>c.
\end{array}
\right.
\end{equation}
Let $\sigma:\Sigma=\{0,\ast,1\}^N\rightarrow\Sigma$ be the shift map:
\begin{equation}
\hspace*{-6mm}
\sigma({\bf s}_\tau)={\bf t}_\tau=(t_1,t_2,\cdots,t_n,\ldots),\, 
\mbox{such that}\,\, t_i=s_{i+1}. 
\end{equation}
We adopt the convention that if, for some $j$, we have $s_j=\ast$, 
then we stop the sequence, that is, the itinerary is a finite string. 
Hence, if $\tau^n(x)\neq c$ for all $n$, then ${\bf s}_\tau(x)\in\{0,1\}^\mathbb{N}$.

%
Figure \ref{greedylazymap} (a) (or (b)) shows
such a  greedy (or lazy) expansion map 
$C_{\beta,1}(x)$ (or $C_{\beta,(\beta-1)^{-1}}(x)$) with a unique 
 point of discontinuity $c=\gamma$ (or $c=\gamma(\beta-1)^{-1}$)
which corresponds exactly to the flaky quantiser 
$Q^f_{[\gamma,\gamma(\beta-1)^{-1}]}(\cdot)$ and also 
has its strongly invariant subinterval $[0,1)$ 
(or $(\beta - 1)^{-1}-1,(\beta - 1)^{-1})$) 
because of its {\it locally eventually onto-mapping}. 
Such a map 
defines the greedy (or lazy) dynamical system, 
defined as $([0,1),C_{\beta,1}(x))$ (or $([(\beta - 1)^{-1}-1,
(\beta - 1)^{-1}),C_{\beta,(\beta - 1)^{-1}}(x))$)  
which is illustrated by the bold lines in Fig.\ref{greedylazymap} (a) 
(or (b)). 
%
The strongly invariant subinterval associated with the $L$-iterated greedy (or lazy) map $C^L_{\beta,1}(x)$ (or $C_{\beta,(\beta - 1)^{-1}}^L(x)$),
corresponds to the  subinterval, defined as
\begin{equation}
I^{\text{invariant}}_{L,C^L_{\beta,1}}(x)=
[\displaystyle{\sum_{i=1}^L b_{i,C^i_{\beta,1}}\gamma^i},
 \displaystyle{\sum_{i=1}^L b_{i,C^i_{\beta,1}}\gamma^i+\gamma^L}),
\label{eq:greedyinvarintsubinterval}
\end{equation}
( or 
\begin{eqnarray}
\hspace*{-6mm}
I^{\text{invariant}}_{L,C^L_{\beta,(\beta-1)^{-1}}}(x)
\hspace*{-3mm}
&=&\hspace*{-3mm}
[\displaystyle{\sum_{i=1}^L b_{i,C^i_{\beta,(\beta-1)^{-1}}}\gamma^i
+\gamma^L((\beta-1)^{-1}-1)},\nonumber\\ 
\hspace*{-9mm}
&&\hspace*{-3mm}
\displaystyle{\sum_{i=1}^L b_{i,C^i_{\beta,(\beta-1)^{-1}}}\gamma^i
+\gamma^L(\beta-1)^{-1}})).
\label{eq:lazyinvarintsubinterval}
\end{eqnarray}
Dajani and Kraaikamp~\cite{Dajani}, however, discussed the 
differences between the error of the greedy expansion of $x$, defined as 
$\gamma^LC^L_{\beta,1}(x)$, and that of the lazy expansion of $x$, 
defined as $\gamma^LC^L_{\beta,(\beta-1)^{-1}}(x)$, 
and concluded that \lq\lq {\it on average for almost all $x$, 
the greedy-convergents, defined as 
$\sum^L_{i=1} b_{i,C^i_{\beta,1}}\gamma^i$, 
approximate $x$ \lq better\lq\, 
than the  lazy-convergents of $x$, defined as  
$\sum^L_{i=1} b_{i,C^i_{\beta,(\beta-1)^{-1}}}\gamma^i$.}" 
%
Furthermore, Daubechies {\it et al.}~\cite{Daubechies2k61} used 
$\widehat{x}^\textit{Daubechies {\it et al.}}_{L,C^L_{\beta,\nu_1^L}}=
\sum^L_{i=1} b_{i,C_{\beta,\nu_1^i}} \gamma^i$, 
 here called the "{\it smallest value of the $L$th stage subinterval 
$I_{L,C_{\beta,\nu_1^L}^L}(x)$}", denoted by $\widehat{x}(\gamma,0)$, 
 and defined by
\begin{equation}
\widehat{x}(\gamma,0)=\displaystyle{\sum_{i=1}^L b_{i,C^i_{\beta,\nu_1^i}}\gamma^i}.
\end{equation}
Such a decoded value, however, works in favour of the greedy expansion 
of a sample $x$, {\it i.e.,}\,$\sum_{i=1}^L b_{i,C^i_{\beta,1}}\gamma^i$, 
which is  equal to Dajani and Karikaamp's greedy convergent~\cite{Dajani}  
if $x$ is uniformly and 
independently distributed over the unit interval $[0,1]$.
 This comes from the fact that
the {\it strongly invariant subinterval} of the 
{\it locally eventually onto-(greedy expansion) map} $C^L_{\beta,1}(x)$, 
$I^{\text{invariant}}_{L,C^L_{\beta,1}}(x)$ 
is  skewed towards the left portion of the interval 
$I_{L,C_{\beta,1}^L}(x)$. 
 
On the other hand, the "{\it largest value of the $L$th stage subinterval 
$I_{L,C_{\beta,\nu_1^L}^L}(x)$}", denoted by $\widehat{x}(\gamma,2)$, 
and defined by
\begin{equation}
\widehat{x}(\gamma,2)=\displaystyle{\sum_{i=1}^L b_{i,C^i_{\beta,\nu_1^i}}\gamma^i+\gamma^L(\beta-1)^{-1}}
\end{equation}
  works in  favour of the \lq\lq {\it steady}" lazy expansion, {\it i.e.,}
$\sum_{i=1}^L b_{i,C^i_{\beta,(\beta-1)^{-1}}}\gamma^i$, 
which is  equal to Dajani and Karikaamp's lazy convergent, 
as shown in 
Fig.\ref{Varxp_L=012}.
That is, the lazy expansion defines the {\it strongly invariant subinterval}  
of the {\it locally eventually onto-(lazy expansion) map} 
$C^L_{\beta,(\beta-1)^{-1}}(x)$, 
$I^{\text{invariant}}_{L,C^L_{\beta,(\beta-1)^{-1}}}(x)$, 
which is  skewed towards the right portion of the interval 
$I_{L,C^L_{\beta,(\beta-1)^{-1}}}(x)$.

Two locally eventually onto R\'enyi maps,  as shown in Fig.3(a) and (b), 
  demonstrate that greedy expansion  and lazy expansion maps are symmetrical 
as follows.
The lazy expansion of $x$ and the greedy expansion of $\psi(x)$, 
respectively  defined by  
\begin{equation}
\begin{array}{lcl}
\hspace*{-4mm}
x\hspace*{-2mm}&=&
\displaystyle{\sum^L_{i=1}} b_{i,C^i_{\beta,(\beta -1)^{-1}}}\gamma^i 
+\gamma^LC_{\beta,(\beta -1)^{-1}}^L(x),\\ 
\hspace*{-4mm}
\psi(x)\hspace*{-2mm}&=&
\displaystyle{\sum^L_{i=1}} c_{i,C^i_{\beta,1}}\gamma^i 
+\gamma^LC_{\beta,1}^L(\psi(x)),
\end{array}
\label{eq.greedylazy}
\end{equation}
satisfy the following lemma~\cite{Dajani}:

{\bf Lemma $1$}:\,{\it The greedy map $C_{\beta,1}^L(x)$ and the lazy map
$C_{\beta,(\beta -1)^{-1}}^L(x)$ are {\it topologically conjugate}, 
{\it i.e.,} 
\begin{equation}
\psi(C_{\beta,(\beta -1)^{-1}}^L(x))= 
C_{\beta,1}^L(\psi(x)),L\in\mathbb{Z}.
\end{equation}
}

{\itshape proof:}
Using Eq.(\ref{eq.greedylazy}) and 
$c_{i,C^i_{\beta,1}}=\overline{b}_{i,C^i_{\beta,(\beta -1)^{-1}}}$, we get
\begin{eqnarray}
x+ \psi(x)\hspace*{-2mm}&=&\hspace*{-4mm} 
\sum^L_{i=1} (b_{i,C^i_{\beta,(\beta -1)^{-1}}}
             +c_{i,C^i_{\beta,1}})\gamma^i
\nonumber\\
\hspace*{-2mm}&+&\hspace*{-2mm}\gamma^L
\{C_{\beta,(\beta-1)^{-1}}^L(x)+C_{\beta,1}^L(\psi(x))\}\nonumber\\
\hspace*{-2mm}&=&\hspace*{-2mm}\sum^L_{i=1} \gamma^i
+\gamma^L\{C_{\beta,(\beta -1)^{-1}}^L(x)+C_{\beta,1}^L(\psi(x))\}\nonumber\\
\hspace*{-2mm}&=&\hspace*{-2mm}
\displaystyle{\frac{\gamma-\gamma^{L+1}}{1-\gamma}}
+\gamma^L\{C_{\beta,(\beta -1)^{-1}}^L(x)+C_{\beta,1}^L(\psi(x))\}.\nonumber
\end{eqnarray}
 The trivial relation 
$x+ \psi(x)=(\beta-1)^{-1}=\displaystyle{\frac{\gamma}{1-\gamma}}$ 
gives
\begin{equation}
\hspace*{-3mm}
 (\beta - 1)^{-1}=C_{\beta,(\beta -1)^{-1}}^L(x)+C_{\beta,1}^L(\psi(x)),\,
L\in\mathbb{Z}
 \label{eq.rel1}
\end{equation}
or 
\begin{equation}
 \psi(C_{\beta,(\beta -1)^{-1}}^L(x))=
C_{\beta,1}^L(\psi(x))
 \label{eq.topologicalgreedylazy}
\end{equation}
which completes the proof.\hfill $\Box$

This lemma implies that 
many dynamical properties of the greedy dynamical system  are preserved by the 
conjugacy;  that is, 
 topologically conjugate systems are dynamically the same in this sense. 
\footnote{Let $\mu_{\beta,1}$ 
be the greedy measure whose density is given by $\rho(x)$ for $\alpha=0$, 
{\it i.e.,} $d\mu_{\beta,1}\propto\rho(x)dx$ 
and $\mu_{\beta,(\beta-1)^{-1}}$ 
the lazy one, 
then for any Lesbesgue set $A\subset [0,(\beta-1)^{-1}]$, the relation 
$\mu_{\beta,(\beta-1)^{-1}}(A)=\mu_{\beta,1}(\psi^{-1}(A))$ 
holds~\cite{Dajani}.}
These two expansions satisfy the following strong relation
which provides 
a starting point for this study:

{\bf Theorem $1$}:\,\,
{\it
Let $\widehat{x}_{L,C^L_{\beta,(\beta - 1)^{-1}}}$ 
be the decoded value of $x$ using its lazy 
expansion $\{b_{i,C^i_{\beta,(\beta - 1)^{-1}}}\}_{i=1}^L$, 
defined by
\begin{equation}
\widehat{x}_{L,C^L_{\beta,(\beta - 1)^{-1}}}= \sum^L_{i=1} b_{i,C^i_{\beta,(\beta - 1)^{-1}}} \gamma^i + \dfrac{(\beta-1)^{-1} \gamma^L}{2}.
\label{eq:greedyex}
\end{equation}
Let $\widehat{\psi(x)}_{L,C^L_{\beta,1}}$ be the decoded value of 
$\psi(x)$ using its greedy expansion $\{c_{i,C^i_{\beta,1}}\}_{i=1}^L$, 
defined by 
\begin{equation}
\hspace*{-4mm}
\widehat{\psi(x)}_{L,C^L_{\beta,1}}= \sum^L_{i=1} 
c_{i,C^i_{\beta,1}} \gamma^i + \dfrac{(\beta-1)^{-1} \gamma^L}{2}.
\label{eq:lazyex}
\end{equation}
Then
\begin{equation}
x - \widehat{x}_{L,C^L_{\beta,(\beta -1)^{-1}}}=
\widehat{\psi(x)}_{L,C^L_{\beta,1}}- \psi(x).
\end{equation}
}
{\itshape proof:}
Equations (\ref{eq:greedyex}) and (\ref{eq:lazyex}) immediately yield
\begin{eqnarray}
\hspace*{-10mm}&&
x -\widehat{x}_{L,C^L_{\beta,(\beta -1)^{-1}}}
=\gamma^L\hspace*{-1mm}\left(
C_{\beta,(\beta -1)^{-1}}^L(x)-\dfrac{(\beta-1)^{-1}}{2}\hspace*{-1mm}\right)
,\\
\hspace*{-10mm}&&
\widehat{\psi(x)}_{L,C^L_{\beta,1}}-\psi(x)
\nonumber\\
\hspace*{-10mm}&&
=\gamma^L\left(\dfrac{(\beta-1)^{-1}}{2}-C_{\beta,1}^L(\psi(x))
\right)
\label{eq.rel2}
\end{eqnarray}
which together with Eq.(\ref{eq.topologicalgreedylazy}) completes the proof.
\hfill $\Box$

 The invariant subinterval of the $L$-iterated cautious map 
$C^L_{\beta,\nu^*}(x),\, \nu^*\in(1,(\beta-1)^{-1})$ 
is defined as 
\begin{equation}
\hspace*{-6mm}
I^{\text{invariant}}_{L,C^L_{\beta,\nu^*}}(x)=
[\displaystyle{\sum_{i=1}^L b_{i,C^i_{\beta,\nu^*}}\gamma^i
+\gamma^L(\nu^*-1)},
  \displaystyle{\sum_{i=1}^L b_{i,C^i_{\beta,\nu^*}}\gamma^i
+\gamma^L\nu^*}) 
\label{eq:cautious expansion interval}
\end{equation}
which has as special cases
$I^{\text{invariant}}_{L,C^L_{\beta,1}}(x)$ and 
$I^{\text{invariant}}_{L,C^L_{\beta,(\beta-1)^{-1}}}(x)$. 
It is noteworthy that such a map restricted to the invariant subinterval 
with its {\it discontinuous point} $c=\gamma\nu$ 
 is the same as Parry's $(\beta,\alpha)$-map with its point of discontinuity 
 $c=\gamma(1-\alpha)$~\cite{Parry64}, 
as shown in Fig.\ref{cautiousmap}(b), if $\alpha=\beta(\nu-1)$. 

Throughout this paper, we assume that the forward orbit of 
a point of discontinuity $c=\gamma\nu,\,1\leq\nu\leq(\beta-1)^{-1}$ 
under the $\beta$-expansion map $C_{\beta,\nu}(x)$ is infinite and 
that $c$ is not {\it attracted} to a periodic orbit, 
meaning that there does not exist an $n$-periodic point $x$ 
such that $\lim_{k\rightarrow\infty}f^{kn}(c)=x$. 
Then, the binary sequence $\{b_{i,C^i_{\beta,\nu}}\}_{i=1}^\infty$ governed by 
$C_{\beta,\nu}(x)$ is exactly the itinerary of $x$ under $C_{\beta,\nu}(x)$. 

{\bf Remark $1$}: Let $\tau(\cdot)$ be the greedy, lazy or cautious 
dynamical system, 
defined as $([0,1),C_{\beta,1}(x))$, 
$([(\beta - 1)^{-1}-1,(\beta - 1)^{-1}),C_{\beta,(\beta - 1)^{-1}}(x))$) or 
$([\nu-1,\nu),C_{\beta,\nu}(x))$, respectively
and $M(x)\in\mathbb{N}$ 
the iteration number of a real number $x\in(0,1]$, 
called the {\it first visit time to $J$ of $x$}, 
such that $\tau^n(x)\not\in J, \,n<M,\,\tau^M(x)\in J$, 
where $J$ is given by $[0,1)$, 
$[(\beta - 1)^{-1}-1,(\beta - 1)^{-1})$ or $[\nu-1,\nu)$, respectively. 
Then $M(x)$ is a random variable 
depending on $x$ but if $x\in J$, then $M(x)=0$. 
\section{Interval Partition  by $\beta$-map with varying $\nu$}
Equation (\ref{eq:cautious expansion}), 
 substituting $m-1$ by $L$ as well as 
Eq.(\ref{eq:greedy expansion}) 
(or (\ref{eq:lazy expansion})) as special cases, 
shows that $x$ can be decomposed into two terms, the principal term 
at $L$ bit precision and the residue term $\gamma^{L}C_{\beta,\nu^*}^{L}(x)$.
This enables us to make the elementary observation that $x$ is always confined 
to the contracted subinterval, defined as 
\begin{equation}
\hspace*{-4mm}
I_{L,C^{L}_{\beta,\nu^*}}(x)=
[\displaystyle{\sum_{i=1}^{L}b_{i,C^i_{\beta,\nu^*}}\gamma^i},\displaystyle{\sum_{i=1}^{L}b_{i,C^i_{\beta,\nu^*}}\gamma^i}+\gamma^{L}(\beta-1)^{-1})
\label{eq:cautiousinvarintsubinterval}
\end{equation} 
since $0\leq C_{\beta,\nu^*}^{L}(x)\leq(\beta-1)^{-1}$.
Both this decomposition of $x$ and  the contracted subinterval are obtained 
under the assumption of fixed $\nu^*$ and the {\it onto-mapping property} 
of the above three dynamical systems. 

We are now ready to  study the contraction process of the interval 
with varying $\nu_i\in [1, (\beta - 1)^{-1}],\,1\leq i\leq L$. 
Consider a \lq\lq dynamical" version of Daubechies {\it et al.}'s 
proof~\cite{Daubechies2k2,Daubechies2k61,Daubechies2k62} 
for the exponential accuracy of the $\beta$-encoder containing 
 the flaky quantiser with its threshold 
 $\nu_i$, 
where at each iteration the value of $\nu_i$ may vary. 
Since the fluctuating $\nu_i$ implies 
that each mapping is a kind of nonlinear \lq\lq time-varying" system, 
we have to examine the binary sequences generated by the map 
and the exponential accuracy of the 
$\beta$-encoder  when the value of $\nu$ may vary 
at each iteration.

Let $I_{i, C^i_{\beta,\nu_1^i}}(x)=[l_i(\nu_1^i),r_i(\nu_1^i)),\,i\geq1$ be 
the  $i\mbox{th}$ interval by the R\'enyi (\lq\lq cautious") map 
$C^i_{\beta,\nu_1^i}(x)\,(1\leq\nu_i\leq(\beta -1)^{-1})$ , recursively defined by
\begin{equation}
\begin{array}{lcr}
l_i(\nu_1^i)&=&l_{i-1}(\nu_1^{i-1})+b_{i,C^i_{\beta,\nu_1^i}}\gamma^i,\\
r_i(\nu_1^i)&=&r_{i-1}(\nu_1^{i-1})-\overline{b}_{i,C^i_{\beta,\nu_1^i}}\gamma^i\end{array}
\label{eq:recursionfomula}
\end{equation}
together with the initial interval 
$I_{0,C_{\beta,\nu}}(x)=[l_0,r_0)$ with $l_0=0$ and $r_0=(\beta-1)^{-1}$.
This yields the relations 
\begin{equation}
\begin{array}{lcr}
l_i(\nu_1^i)&=&\displaystyle{\sum_{j=1}^ib_{j,C^j_{\beta,\nu_1^j}}\gamma^j},\\
r_i(\nu_1^i)&=&\displaystyle{(\beta-1)^{-1}-\sum_{j=1}^i\overline{b}_{j,C^j_{\beta,\nu_1^j}}\gamma^j}.
\end{array}
\end{equation}
Using the two trivial relations 
$(\beta-1)^{-1}
=\sum_{j=1}^\infty\gamma^j,\,
(\beta-1)^{-1}-\sum_{j=1}^i\gamma^j
=\sum_{j=i+1}^\infty\gamma^j
=(\beta-1)^{-1}\gamma^i$
 and we can rewrite $r_i(\nu_1^i)$ as
\begin{equation}
\hspace*{-6mm}
r_i(\nu_1^i)=l_i(\nu_1^i)+\displaystyle{\sum_{j=i+1}^\infty\gamma^j}
=l_i(\nu_1^i)+(\beta-1)^{-1}\gamma^i.
\end{equation}
Then, we obtain the useful relation
\begin{equation}
\displaystyle{\frac{|I_{i+1,C^{i+1}_{\beta,\nu_1^{i+1}}}(x)|}{|I_{i,C^i_{\beta\nu_1^i}}(x)|}}
=\gamma,
\label{eq:reductionratio}
\end{equation}
where $|I|$ denotes the width of an interval $I$.
\begin{figure}[htbp]
\begin{center}
\includegraphics[scale=0.3]{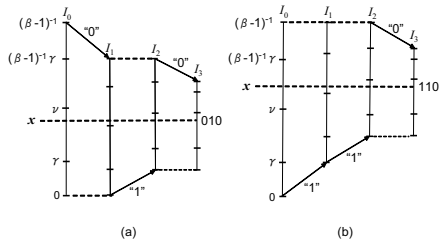} 
\caption{Representation of the $\beta$-expansion process:
the vertical bar with a scale represents the subinterval $I_i(\cdot)$ 
at the $i$th stage, where $I_0=[0,(\beta-1)^{-1})$ is the initial interval. 
A succession of three binary decisions using the quantiser 
$Q^f_\nu(\cdot)$ gives two binary expansions of the sample $x$:(a) $010$ and 
(b) $110$, each of which depends on $\nu$. The widths of the subintervals are 
contracted by $\gamma$ and renormalized.}
\label{PartitionProcessBeta}
\end{center}
\end{figure}
 Figures \ref{PartitionProcessBeta}(a) and (b) 
show two examples of the interval contraction process by $\beta$-expansion,  
where the subinterval is 
marked with a scale that indicates several numbers and a renormalization rule 
is devised which is guaranteed 
by the onto-mapping $C_{\beta,\nu}(x)$. 
Furthermore, this yields the following important lemma.

{\bf Lemma $2$}:{\it 
If $x\in I_{i,C^i_{\beta,\nu_1^i}}(x)$, then $x\in 
I_{i+1,C^{i+1}_{\beta,\nu_1^{i+1}}}(x),\,\, i\in\mathbb{Z}$. That is, 
 $x$ is always confined to the $i$th subinterval where the binary digits 
of the $\beta$-expansion of $x$ are obtained.
}

{\itshape proof:}
It is obvious that $x\in I_{0,C_{\beta,\nu}}(x)=[0,(\beta-1)^{-1})$ because 
$x\in[0,1)\subset I_{0,C_{\beta,\nu}}(x)$.
Suppose that $x\in I_{i,C^i_{\beta,\nu_1^i}}(x),\,i\geq1$.
\begin{eqnarray*}
\hspace*{-4mm}
&&\hspace*{-2mm}x-l_{i+1}(\nu_1^{i+1})=
\displaystyle{\sum_{j=1}^ib_{j,C^j_{\beta,\nu_1^j}}\gamma^j}+\gamma^i
C^i_{\beta,\nu_1^i}(x)\\
\hspace*{-4mm}&-&\hspace*{-2mm}
\{l_i(\nu_1^i)+b_{i+1,C^{i+1}_{\beta,\nu_1^{i+1}}}\gamma^{i+1}\}
=\gamma^i\{C^i_{\beta,\nu_1^i}(x)-\gamma b_{i+1,C^{i+1}_{\beta,\nu_1^{i+1}}}\},
\end{eqnarray*}
which implies that if 
$C_{\beta,\nu_1^{i}}^i(x)<\gamma\nu_{i+1},
\mbox{{\it i.e.,}} \, b_{i+1,C^{i+1}_{\beta,\nu_1^{i+1}}}=0$, then 
$x-l_{i+1}(\nu_1^{i+1})=\gamma^iC^i_{\beta,\nu_1^i}(x)\geq0$.
Otherwise, {\it i.e.,}\, if
$C^i_{\beta,\nu_1^{i}}(x)\geq\gamma
\nu_{i+1},\,b_{i+1,C^{i+1}_{\beta,\nu_1^{i+1}}}=1$, 
then $x-l_{i+1}(\nu_1^{i+1})
\geq\gamma^{i+1}(\nu_{i+1}-1)\geq0$ 
since $1\leq\nu_{i+1}$.
On the other hand, 
\begin{eqnarray*}
\hspace*{-6mm}
&&\hspace*{-3mm}
r_{i+1}(\nu_1^{i+1})-x\\
\hspace*{-6mm}
&=&\hspace*{-3mm}(\beta-1)^{-1}
-\displaystyle{\sum_{j=1}^{i+1}\overline{b}_{j,C^j_{\beta,\nu_1^{j}}}
\gamma^{j}}
-\{\displaystyle{\sum_{j=1}^ib_{j,C^j_{\beta,\nu_1^j}}\gamma^j}
+\gamma^iC^i_{\beta,\nu_1^i}(x)\}\\
\hspace*{-6mm}
&=&\hspace*{-3mm}\displaystyle{\sum_{j=1}^{\infty}\gamma^j-\sum_{j=1}^{i+1}\gamma^j}
+b_{i+1,C^{i+1}_{\beta,\nu_1^{i+1}}}\gamma^{i+1}
-\gamma^{i}C^i_{\beta,\nu_1^i}(x)\\
\hspace*{-6mm}
&=&\hspace*{-3mm}\gamma^{i+1}\{(\beta-1)^{-1}+b_{i+1,C^{i+1}_{\beta,\nu_1^{i+1}}}
-\beta C^i_{\beta,\nu_1^i}(x)\},
\end{eqnarray*}
which implies that if 
$C_{\beta,\nu_1^{i}}^i(x)<\gamma\nu_{i+1},
\mbox{{\it i.e.,}}\, b_{i+1,C^{i+1}_{\beta,\nu_1^{i+1}}}=0$, then 
$r_{i+1}(\nu_1^{i+1})-x=\gamma^{i+1}\{(\beta-1)^{-1}-
\beta C^i_{\beta,\nu_1^i}(x)\}
\geq\gamma^{i+1}\{(\beta-1)^{-1}-\nu_{i+1}\}\geq0$ 
because $\nu_{i+1}\leq(\beta-1)^{-1}$.
Otherwise, {\it i.e.,}\, if
$C^i_{\beta,\nu_1^{i}}(x)\geq
\gamma\nu_{i+1},\,b_{i+1,C^{i+1}_{\beta,\nu_1^{i+1}}}=1$, 
then $r_{i+1}(\nu_1^{i+1})-x=\gamma^{i+1}\{(\beta-1)^{-1}+1-\beta 
C^i_{\beta,\nu_1^i}(x)\}
\geq0$ 
since $C^i_{\beta,\nu_1^i}(x)\leq(\beta-1)^{-1}$.
This implies that $x\in I_{i+1,C^{i+1}_{\beta,\nu_1^{i+1}}}(x)$. 
This completes the proof.\hfill $\Box$
\begin{figure}[htbp]
\begin{center}
\includegraphics[scale=0.45]{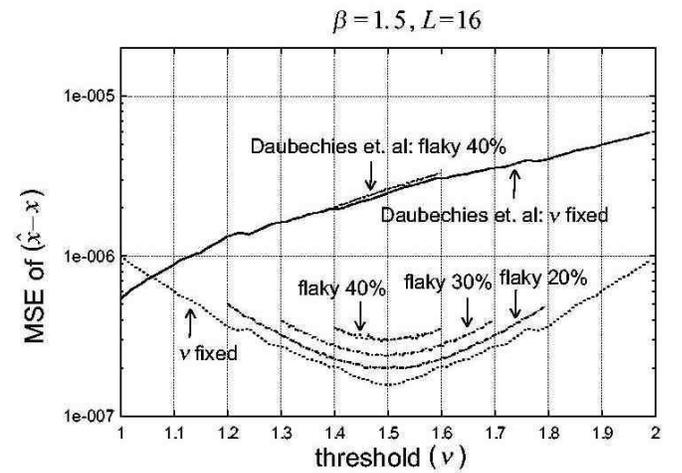} 
\end{center}
\caption{The 
$\mbox{MSE}
(\widehat{x}_{L,C^L_{\beta,\nu_1^L}}(\gamma,p_L))$, \,$p_L\in\{0,1\}$\, 
using the exact $\beta$ of the $\beta$-encoder with $\beta =1.5$, $L=16$, 
 and  fluctuating $\nu_i,\,1\leq i\leq L$\, with 
its several fluctuation bounds $0,10,20,30,40\%$.
}
\label{Varxflakyp_L1estbeta}
\end{figure}

Thus, we have obtained Eq.(\ref{eq:varying cautious expansion interval}).
%
We differentiate  between the choices of fixed $\nu^*$, {\it e.g.,} the 
greedy/lazy and cautious expansions and the decoded methods of a sample $x$. 
However, the roles of "0" and "1" should be equal in the binary expansion of 
a sample $x$.  
Theorem 1 supports this simple intuition which is an elementary result but is 
a fundamental point for this study. 
Furthermore, Lemma $2$ readily leads us to a {\it good} decoded value of 
$x$, {\it i.e.,}
 the "{\it midpoint of the $L$th stage subinterval $I_{L,C_{\beta,\nu_1^L}^L}(x)$ with $\,\nu_i\in[1,(\beta-1)^{-1}],\,1\leq i\leq L$}",
denoted by $\widehat{x}(\gamma,1)$, and defined by
\begin{equation}
\widehat{x}(\gamma,1)=\displaystyle{\sum_{i=1}^L b_{i,C_{\beta,\nu_1^i}}\gamma^i+\frac{\gamma^L(\beta-1)^{-1}}{2}}.
\end{equation}
This works in favour of the cautious expansion with its 
{\it strongly invariant subinterval of the locally eventually 
onto-(cautious expansion) map} 
$C^L_{\beta,\nu^*}(x)$, 
$I^{\text{invariant}}_{L,C^L_{\beta,\nu^*}}(x)$.
%

Hence we readily obtain the following important theorem: 

{\bf Theorem $2$}~\cite{NDES2007}:\,{\it 
If we introduce different  decoded values of 
$x$, depending on the representative point in the $L$th subinterval, 
denoted by the index $p_L$ as follows:
\begin{equation}
\widehat{x}_{L,C^L_{\beta,\nu_1^L}}(\gamma,p_L)=
\left\{
\begin{array}{lr}
l_L(\nu_1^L),
&\mbox{if $p_L=0$},\\
\displaystyle{\frac{l_L(\nu_1^L)+r_L(\nu_1^L)}{2}},
&\mbox{if $p_L=1$},\\
r_L(\nu_1^L),
&\mbox{if $p_L=2$},
\end{array}
\right.
\label{eq:cautious decoded for x with p_L}
\end{equation}
then at $L$-bit precision, the approximation error between the original value $x$ and its decoded value $\widehat{x}_{L,C^L_{\beta,\nu_1^L}}(\gamma,p_L)$ 
is bounded by 
\begin{equation}
|x -\widehat{x}_{L,C^L_{\beta,\nu_1^L}}(\gamma,p_L)| 
\le \frac{(1+|p_L-1|)}{2}\cdot(\beta - 1)^{-1} \gamma^L.
\end{equation}
}
{\it Proof}: 
Since $x \in [l_L(\nu_1^L), r_L(\nu_1^L)]$, the approximation error 
is bounded by 
\begin{equation}
|x - \widehat{x}_{L,C^L_{\beta,\nu_1^L}}(\gamma,p_L)|
\leq 
\left\{
\begin{array}{lr}
(\beta-1)^{-1}\gamma^L,&\mbox{if $p_L=0\,\mbox{or}\,2$},\\
\frac{(\beta-1)^{-1}\gamma^L}{2},&\mbox{if $p_L=1$}.
\end{array}
\right.
\end{equation}
This concludes the proof.\hfill $\Box$\\
{\bf Remark $2$}:
Theorem $2$ demonstrates that 
the decoded value of $x$ using $\{b_{i,C^i_{\beta,\nu_1^i}}\}_{i=1}^L$ 
should be defined by $\widehat{x}_{L,C^L_{\beta,\nu_1^L}}(\gamma,1)$. 
On the other hand, $\widehat{x}_{L,C^L_{\beta,\nu_1^L}}(\gamma,0)$ is 
identical to the decoded value of Daubechies {\it et al.} 
Namely, $\widehat{x}_{L,C^L_{\beta,\nu_1^L}}(\gamma,1)$ 
improves the quantisation error  by $3$dB over the Daubechies {\it et al.}'s bound $\nu\gamma^L$~\cite{Daubechies2k61}  when $\beta>1.5$ since 
\begin{equation}
|x -\widehat{x}_{L,C^L_{\beta,\nu_1^L}}(\gamma,1)| 
\le \frac{(\beta - 1)^{-1} \gamma^L}{2} < \gamma^L\leq\nu\gamma^L.
\end{equation}
These observations  are clearly confirmed in several numerical results 
of the mean square error (MSE) of $\widehat{x}_{L,C^L_{\beta,\nu_1^L}}(\gamma,1)$ 
using $\beta$ as shown in Fig.\ref{Varxp_L=012} and those of estimated 
$\widehat{\beta}$ as shown below in Fig.\ref{Varbeta1} as well as those of 
$\widehat{x}_{L,C^L_{\beta,\nu_1^L}}(\widehat{\gamma},1)$
 using estimated $\widehat{\beta}$ in Fig.\ref{Varxestimatebeta}. 
 Figure \ref{Varxp_L=012} shows the MSE of quantisations 
by decoded 
$\widehat{x}_{L,C^L_{\beta,\nu_1^L}}(\gamma,p_L),\,0\leq p_L\leq2$
 using the value $\beta=1.5$ of the $\beta$-encoder with  $L=16$ 
for fixed $\nu^*\in[1,(\beta-1)^{-1}]$.
It is clear that
the  MSE of the decoded $\widehat{x}_{L,C^L_{\beta,\nu^*}}$ by the {\it cautious} expansion is smaller than those of the {\it greedy} and {\it lazy} expansions because of their invariant subintervals (see  Fig.\ref{Invariant subinterval}(a)). 

 In all of the numerical simulations as discussed below, 
we average over $10,000$ 
samples $x$, 
which are assumed to be uniformly and independently 
distributed over $[0,1]$ for $100$  thresholds  
$\nu^\ast\in[1,(\beta-1)^{-1}]$  with its associated 
fluctuating thresholds $\nu_i,\,1\leq i\leq L$ 
based on the fluctuations $u_i$\,, {\it i.e.,}
$\nu_i=\nu^\ast(1+u_i)$ 
where $u_i$ is an independent random variable with 
 bound $\varepsilon\in \{0,0.2,0.3,0.4\}$.
We introduce the mean squared  error (MSE) of the decoded 
$\widehat{x}_{L,C^L_{\beta,\nu_1^L}}$, defined as
\begin{equation}
\mbox{MSE}(\widehat{x}_{L,C^L_{\beta,\nu_1^L}})=\displaystyle{\int_0^1
(x-\widehat{x}_{L,C^L_{\beta,\nu_1^L}})^2dx.}
\end{equation}
Figure \ref{Varxflakyp_L1estbeta} shows the 
$\mbox{MSE}(
\widehat{x}_{L,C^L_{\beta,\nu_1^L}}(\gamma,p_L))$,$\,p_L\in\{0,1\}$ 
using the value of $\beta$  for the fluctuating 
$\nu_i,\,1\leq i\leq L$\, with its several fluctuation bounds. 
\begin{figure}[htbp]
\begin{center}
\includegraphics[scale=0.45]{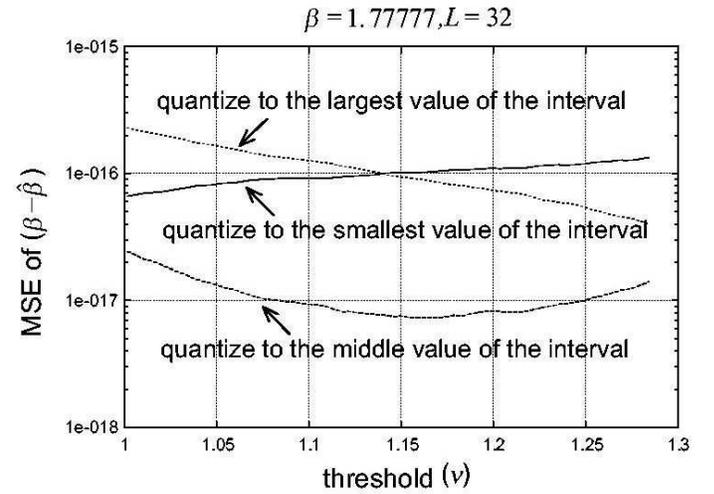} 
\end{center}
\caption{The MSE of the estimated $\beta$, {\it i.e.,} 
$\widehat{\beta}_{L,C^L_{\beta,\nu^*}}(\widehat{x},0)$, 
$\widehat{\beta}_{L,C^L_{\beta,\nu^*}}(\widehat{x},1)$, and 
$\widehat{\beta}_{L,C^L_{\beta,\nu^*}}(\widehat{x},2)$
of the $\beta$-encoder with $\beta =1.77777$, $L=32$, 
and fixed $\nu^*\in [1,(\beta-1)^{-1}]$.
}
\label{Varbeta1}
\end{figure}
\begin{figure}[htbp]
\begin{center}
\includegraphics[scale=0.45]{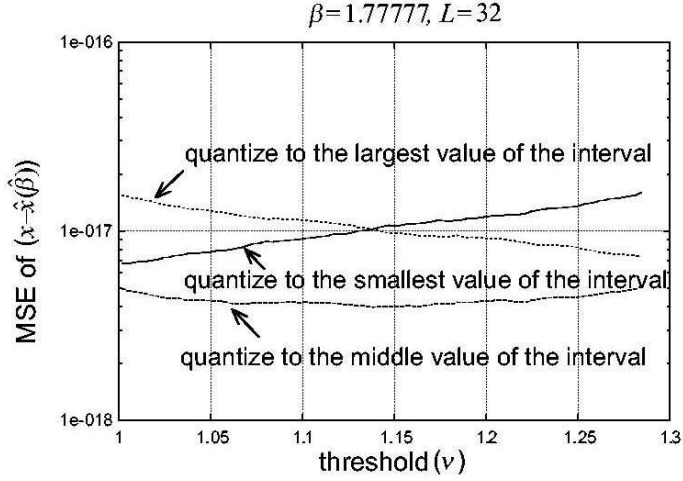} 
\end{center}
\caption{The MSE of quantisations by decoded   
$\widehat{x}_{L,C^L_{\beta,\nu^*}}(\widehat{\gamma},0)$,  
$\widehat{x}_{L,C^L_{\beta,\nu^*}}(\widehat{\gamma},1)$\, and 
$\widehat{x}_{L,C^L_{\beta,\nu^*}}(\widehat{\gamma},2)$
using the estimated $\widehat{\beta}$ of the $\beta$-encoder 
with  $\beta =1.77777$, $L=32$, and fixed 
$\nu^*\in[1,(\beta-1)^{-1}]$.
}
\label{Varxestimatebeta}
\end{figure}
\begin{figure}[htbp]
\begin{center}
\includegraphics[scale=0.45]{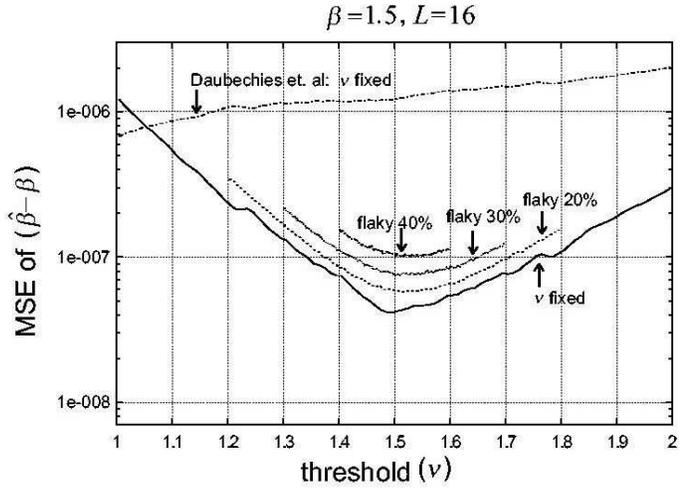} 
\end{center}
\caption{The MSE of the estimated $\beta$, 
{\it i.e.,}
$\widehat{\beta}_{L,C^L_{\beta,\nu_1^L}}(\gamma,1)$ and 
$\widehat{\beta}_{L,C^L_{\beta,\nu_1^L}}(\gamma,0)$\, 
 of the $\beta$-encoder with  $\beta =1.5$, $L=16$, and 
fluctuating $\nu_i,\,1\leq i\leq L$\,
with its fluctuation bounds $0,20,30,40\%$.
}
\label{Varxfp_L2}
\end{figure}
\begin{figure}[htbp]
\begin{center}
\includegraphics[scale=0.45]{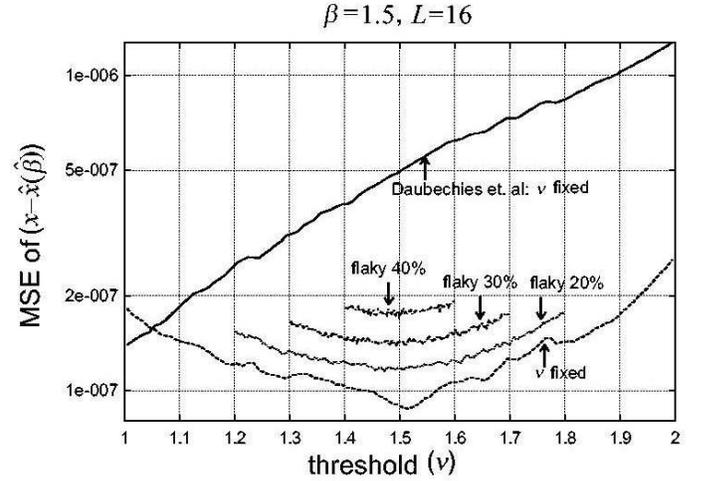} 
\end{center}
\caption{The MSE of quantisations by decoded   
$\widehat{x}_{L,C^L_{\beta,\nu_1^L}}(\widehat{\gamma},0)$\, and
$\widehat{x}_{L,C^L_{\beta,\nu_1^L}}(\widehat{\gamma},1)$\, 
using the estimated $\beta$ {\it i.e.,}    
$\widehat{\beta}_{L,C^L_{\beta,\nu_1^L}}(\gamma,0)$\, and
$\widehat{\beta}_{L,C^L_{\beta,\nu_1^L}}(\gamma,1)$\, 
of the $\beta$-encoder with $\beta =1.5$, $L=16$, and fluctuating 
$\nu_i,\,1\leq i\leq L$ with its fluctuation bounds $0,20,30,40\%$.
}
\label{Varbeta2}
\end{figure}
\section{Characteristic equations for $\beta$}
In order to show the self-correction property of the 
amplification factor $\beta$, Daubechies {\it et al.}~\cite{Daubechies2k62} 
gave an equation for $\beta$ 
governed 
by the bit sequences of the sample data as follows.
Using the $\beta$-expansion sequences $b_{i,C^i_{\beta,\nu_1^i}}$ 
for $x \in [0,1)$ 
and $c_{i,C^i_{\beta,\nu_1^i}}$ for $y = 1 - x$ $(i=1,2,\cdots,L)$ 
yields a root of the 
\textit{characteristic equation of $\beta$}, defined by
\begin{equation}
P^\textit{Daubechies {\it et al.}}_{L,C^L_{\beta,\nu_1^L}}(\gamma) = 
1 - \sum^{L}_{i=1} (b_{i,C^i_{\beta,\nu_1^i}} + c_{i,C^i_{\beta,\nu_1^i}}) 
\gamma^i,
\label{eq:Daubcharacteristic equation}
\end{equation}
as the estimated $\gamma$.
In order to  apply Daubechies {\it et al.}'s idea for estimating $\beta$, 
let us  introduce  cautious expansions $\{b_{i,C^i_{\beta,\nu_1^i}}\}_{i=1}^L$ for $x$ and $\{c_{i,C^i_{\beta,\nu_1^i}}\}_{i=1}^L$ for $y = 1 - x$. 
Let us use Eq.(\ref{eq:cautious decoded for x with p_L})  to define 
the decoded values, respectively as  follows:
\begin{equation}
\begin{array}{lcr}
\widehat{x}_{L,C^L_{\beta,\nu_1^L}}(\gamma,p_L)& =&
\displaystyle{
\sum^L_{i=1} b_{i,C^i_{\beta,\nu_1^i}} \gamma^i + 
p_L\cdot\frac{\gamma^{L+1}}{2(1-\gamma)}}
, \\
\widehat{y}_{L,C^L_{\beta,\nu_1^L}}(\gamma,p_L)& =& 
\displaystyle{
\sum^L_{i=1} c_{i,C^i_{\beta,\nu_1^i}} \gamma^i + 
p_L\cdot\frac{\gamma^{L+1}}{2(1-\gamma)}}.
\end{array}
\end{equation}
Then we get the relation 
\begin{eqnarray}
\hspace*{-6mm}
&&\widehat{x}_{L,C^L_{\beta,\nu_1^L}}(\gamma,p_L) + 
\widehat{y}_{L,C^L_{\beta,\nu_1^L}}(\gamma,p_L) \nonumber\\
\hspace*{-6mm}
&&=
\sum^L_{i=1}\left(b_{i,C^i_{\beta,\nu_1^L}}+c_{i,C^i_{\beta,\nu_1^L}}\right)
\gamma^i
+p_L\cdot\displaystyle{\frac{\gamma^{L+1}}{1-\gamma}}\simeq1
\end{eqnarray}
which gives a new characteristic equation for $\beta$: 

{\bf Theorem $3$}~\cite{NDES2007}:{\it 
The estimated value of $\gamma$ is a root of the polynomial 
$P_{L,C^L_{\beta,\nu_1^L}}(\gamma,p_L)$,  defined by 
\begin{equation}
P_{L,C^L_{\beta,\nu_1^L}}(\gamma,p_L) =1- \sum^{L}_{i=1} (b_{i,C^i_{\beta,\nu_1^i}}+c_{i,C^i_{\beta,\nu_1^i}})\gamma^i-p_L\cdot\frac{\gamma^{L+1}}{1-\gamma}.
\label{eq:betacharacteristicequation}
\end{equation}
}

The uniqueness of such a root of the continuous function 
$P_{L,C^L_{\beta,\nu_1^L}}(\gamma,p_L)$ over the interval $[0,1]$
is guaranteed by the intermediate value theorem 
since 
$\displaystyle{\frac{d\,P_{L,C_{\beta,\nu_1^L}}(\gamma,p_L)}{d\,\gamma}}<0$.
Let $\widehat{\beta}_{L,C^L_{\beta,\nu_1^L}}(\widehat{x})$ be the root of 
$P_{L,C^L_{\beta,\nu_1^L}}(\gamma,p_L)$ as a function of the sample $x$, 
which is uniformly and independently distributed over $[0,1]$. 
We introduce the MSE of the estimated 
$\widehat{\beta}_{L,C^L_{\beta,\nu_1^L}}$, defined as
\begin{equation}
\mbox{MSE}(\widehat{\beta}_{L,C^L_{\beta,\nu_1^L}}(\widehat{x}))=\displaystyle{\int_0^1
(\beta-\widehat{\beta}_{L,C^L_{\beta,\nu_1^L}})^2dx.}
\end{equation}
{\bf Remark $3$}:\, Daubechies {\it et al.}'s 
characteristic equation of $\beta$, 
 Eq.(\ref{eq:Daubcharacteristic equation})~\cite{Daubechies2k62}, {\it i.e.,} 
$P^\textit{Daubechies {\it et al.}}_{L,C^L_{\beta,\nu_1^L}}(\gamma,0)$
has no  term 
$\gamma^{L+1}(1-\gamma)^{-1}$ that can be written as the sum of two  terms 
$\gamma^{L+1}[2(1-\gamma)]^{-1}$ of the decoded values 
$\widehat{x}_{L,C^L_{\beta,\nu_1^L}}$ and 
$\widehat{y}_{L,C^L_{\beta,\nu_1^L}}$, which 
come from $\gamma^LC_{\beta,\nu_1^L}^L(x)$ and 
$\gamma^LC_{\beta,\nu_1^L}^L(1-x)$, respectively. 
However, this missing term $\gamma^{L+1}[2(1-\gamma)]^{-1}$ 
plays an important role in estimating both $\beta$ and $x$  
precisely, as shown below. 
Thus, this term should not be removed because information will be lost. 
This is one of the main differences between Daubechies {\it et al.}'s 
DA conversion in the $\beta$-encoder 
and ours defined here.\\
{\bf Remark $4$}:\,In a decoding process, knowing the exact value of a fixed 
$\nu^*\in[1,(\beta-1)^{-1})$ is unnecessary. 
If one wants to know the estimated $\nu^*$, it is given by  
\begin{equation}
\widehat{\nu}^*_{L,C^L_{\beta,\nu^*}}(\gamma,1)=
\sum^L_{i=1} \overline{b}_{i,C^i_{\beta,\nu^*}}
\gamma^i+\frac{(\beta-1)^{-1}\gamma^L}{2},
\end{equation}
which comes from exchanging the roles of $x$ and $\nu$.

Figures \ref{Varbeta1} and \ref{Varxestimatebeta} show the 
$\mbox{MSE}(\widehat{\beta}_{L,C^L_{\beta,\nu_1^L}}(\widehat{x},p_L))$ 
and $\mbox{MSE}(\widehat{x}_{L,C^L_{\beta,\nu_1^L}}(\widehat{\gamma},p_L))$,  
respectively, for $\,0\leq p_L\leq2$, using the estimated $\widehat{\beta}$ 
of the $\beta$-encoder for fixed $\nu^*$. 
Comparing Fig.\ref{Varxp_L=012} with Fig.\ref{Varxestimatebeta} 
leads us to observe that 
$\widehat{x}_{L,C^L_{\beta,\nu_1^L}}(\widehat{\gamma},p_L)$ gives a better 
approximation to $x$ than $\widehat{x}_{L,C^L_{\beta,\nu_1^L}}(\gamma,p_L)$  
in terms of the MSE performance. 
Figures \ref{Varxfp_L2} and \ref{Varbeta2} show the 
$\mbox{MSE}(\widehat{\beta}_{L,C^L_{\beta,\nu_1^L}}(\widehat{x},p_L))$ and 
$\mbox{MSE}(\widehat{x}_{L,C^L_{\beta,\nu_1^L}}(\widehat{\gamma},p_L))$, 
respectively, with $\,p_L\in\{0,1\}$ using the estimated $\widehat{\beta}$ 
of the $\beta$-encoder with fluctuating $\nu_i,\,1\leq i\leq L$\,
with its several fluctuation bounds. 
 As shown in Figs. \ref{Varxflakyp_L1estbeta} and \ref{Varbeta2},  %
$\widehat{x}_{L,C^L_{\beta,\nu_1^L}}
(\widehat{\gamma},p_L),\,p_L\in\{0,1\}$ 
also gives a better approximation to
 $x$ than $\widehat{x}_{L,C^L_{\beta,\nu_1^L}}(\gamma,p_L)$ 
even in the case of fluctuating $\nu_i$. 

\section{Optimal design of an amplifier\\ with a scale-adjusted map}
\begin{figure}[htbp]
\begin{center}
\includegraphics[scale=0.30]{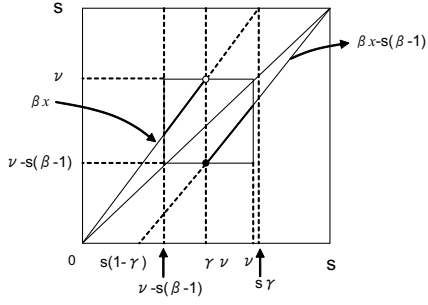} 
\end{center}
\caption{The scale-adjusted $\beta$-map $S_{\beta, \nu, s}(x)$ with its invariant subinterval $[\nu-s(\beta-1),\nu)$.}
\label{SAB}
\end{figure}

\begin{figure}[t]
\begin{center}
\includegraphics[scale=0.30]{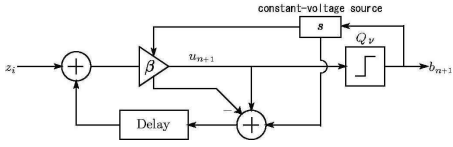} 
\end{center}
\caption{The scale-adjusted $\beta$-encoder, where $s=s(b)$ is defined by 
Eq.(\ref{eqn:s(b)}).}
\label{SABE}
\end{figure}

In $\beta$-encoders, for a given  quantiser tolerance 
$\sigma \le (\beta - 1)^{-1} - 1$, 
we must choose appropriate values for the quantiser threshold $\nu$ 
and the amplifier parameter $\beta$.
The scale of the map  depends solely on $\beta$ which 
determines the MSE of the quantisation. 
This motivates us to introduce a new map, 
called a {\it scale-adjusted map},  with a scale $s>0$ 
 independent of $\beta$, defined by 
\begin{equation}
S_{\beta, \nu, s}(x) =  \left\{
\begin{array}{ll}
\hspace*{-2mm}\beta x, \hspace*{-2mm}&\hspace*{-2mm} x \in [0, \gamma \nu), \\
\hspace*{-2mm}\beta x - s(\beta - 1), \hspace*{-1mm}&\hspace*{-2mm} x \in[\gamma \nu, s), 
\end{array}
\right.\hspace*{-2mm}\nu\in[s(\beta - 1),s),
\label{sbe}
\end{equation}
which is illustrated in Fig. \ref{SAB}. 
This is identical to the  $\beta$-map when $s=(\beta-1)^{-1}$.
Let $b_{i,S^i_{\beta,\nu_1^i,s}}$ be the associated bit sequence for 
the threshold sequence $\nu_1^L$, defined by
\begin{equation}
b_{i,S^i_{\beta,\nu_1^i,s}} = 
\left\{
\begin{array}{ll}
0,       & S_{\beta, \nu_1^{i-1}, s}^{i-1}(x) \in [0, \gamma \nu_i), \\
1, & S_{\beta, \nu_1^{i-1}, s}^{i-1}(x) \in [\gamma \nu_i, s).\\
\end{array}
\right.
\end{equation}
The scale-adjusted map $S_{\beta, \nu, s}(x)$ also determines 
the flaky quantiser $Q^f_{[s(1-\gamma),s\gamma]}(\cdot)$.
The invariant subinterval of $S_{\beta, \nu, s}(x)$, defined as
$[\nu - s(\beta - 1), s\nu)$ is similar to that of the $\beta$-expansions.
Let  $\{b_{i,S^i_{\beta,\nu_1^i,s}}\}_{i=1}^L$ be the sequence generated by 
iterating the map $S_{\beta, \nu, s}(x)$ of $x \in [0, s)$ $L$ times, 
denoted by $S_{\beta, \nu_1^L, s}^L(x)$. Then we get 
\begin{eqnarray}
\hspace*{-6mm}S_{\beta, \nu_1^L, s}^L(x) &=&
\beta S_{\beta, \nu_1^{L-1}, s}^{L-1}(x) - 
s(\beta-1)b_{L,S_{\beta,\nu_1^L,s}}\nonumber \\
&=& \beta^L x - s(\beta - 1)\sum^{L}_{i=1} 
b_{i,S_{\beta,\nu_1^i,s}} \beta^{L-i},
\end{eqnarray}
or 
\begin{equation}
x = s(\beta - 1)\sum^{L}_{i=1} b_{i,S^i_{\beta,\nu_1^i,s}} \gamma^i + 
\gamma^L S_{\beta, \nu_1^L,s}^L(x).
\end{equation}
Using the relation $S^L_{\beta, \nu_1^L, s}(x) \in [0, s)$ gives 
its subinterval, defined  by  
\begin{eqnarray}
\hspace*{-9mm}
&&\hspace*{-4mm}I_{L, S^L_{\beta,\nu_1^L,s}}(x) \nonumber\\
\hspace*{-9mm}
& = &\hspace*{-4mm}
s(\beta-1)[\sum^{L}_{i=1}
b_{i,S^i_{\beta,\nu_1^i,s}} \gamma^i, 
\sum^{L}_{i=1}b_{i,S^i_{\beta,\nu_1^i,s}}
\gamma^i+\displaystyle{\frac{\gamma^{L+1}}{1-\gamma}})
\label{eq:smapinterval}
\end{eqnarray}
which  enables us to easily obtain 
the decoded value $\widehat{x}_{L,S^L_{\beta,\nu_1^L,s}}$ as follows:
\begin{equation}
\widehat{x}_{L,S^L_{\beta,\nu_1^L,s}} = 
s(\beta - 1)\sum^{L}_{i=1} b_{i,S^i_{\beta,\nu_1^i,s}} \gamma^i + 
\dfrac{s \gamma^L}{2}
\end{equation}
and its quantisation error bound as 
\begin{equation}
\epsilon_{L,S_{\beta,\nu_1^L,s}}(x) = 
|x -\widehat{x}_{L,S^L_{\beta,\nu_1^L,s}}|\le \dfrac{s \gamma^L}{2}.
\end{equation}
We introduce an A/D converter, called a \textit{scale-adjusted $\beta$-encoder}, that realizes the above expansion, 
 as shown in Fig. \ref{SABE},
where the scale $s$ of $S_{\beta, \nu, s}$ can be adjusted 
with  the  bit-controlled constant-current source of the quantiser $s(b)$, defined by
\begin{equation}
s(b) = \left\{
\begin{array}{ll}
0,      & b = 0,\\
s,      & b = 1.
\end{array}
\label{eqn:s(b)}
\right.
\end{equation}
Its robustness to the fluctuation of the quantiser threshold $\nu$
is restricted by its tolerance $\sigma_{\beta, s}$: 
\begin{equation}
\sigma_{\beta, s} = s - s(\beta-1)=s(2 - \beta).
\end{equation}
Even if the amplification factor $\beta$ is constant,
the tolerance $\sigma_{\beta, s}$ can be set arbitrarily
by selecting the constant-voltage source $s$.
We get the following convenient lemma as a rule of thumb for AD/DA-converter designs.

{\bf Lemma $3$}:\,{\it 
In a scale adjusted $\beta$-encoder,
for a given bit budget $L$ and quantiser tolerance 
$\sigma_{\beta, s}$, the amplification factor $\beta$
with its inevitable fluctuation, should be set to 
\begin{equation}
\beta = \dfrac{2L}{L+1}.
 \label{eq:bitbudget}
\end{equation}
in order to minimize the quantisation error. 
}

{\itshape proof:}
Eq.(\ref{eq:smapinterval}) gives
$|I_{L, S^L_{\beta, \nu,s}}(x)|=s \beta^{-L} = 
\dfrac{\sigma_{\beta, s} \beta^{-L}}{(2-\beta)}$.
Differentiating $|I_{L, S^L_{\beta, \nu,s}}(x)|$ with respect to $\beta$,
we obtain
\begin{equation}
\dfrac{d|I_{L,S^L_{\beta, \nu,s}}(x)|}{d \beta} = 
\dfrac{\sigma_{\beta, s} \beta^{-L-1}}{(2-\beta)^{2}}\{\beta - L(2-\beta)\}.
\end{equation}
This completes the proof. 
\hfill $\Box$

This lemma shows that the amplification factor $\beta$ minimizing 
the quantisation error is not equal to $2$ 
but the relation $\beta=2$ as in PCM holds only as $L\rightarrow\infty$.
However, we must take a little care at this point.
Assume that for a given initial value of an A/D converter 
 $x \in [0, 1)$, the scale $s=\sigma_{\beta, s}(L+1)/2>1$.
In this case, we should adjust the amplification factor and 
the constant-voltage source 
which satisfy  $\beta = 2L/(L+1)$ and 
$s = \sigma_{\beta, s}(L+1)/2$, respectively; 
otherwise we should adjust these parameters such as $s=1$ and 
$\beta = 2-\dfrac{s}{\sigma_{\beta, s}} = 2-\sigma_{\beta, s}^{-1}$.
\section{Negative $\beta$-encoder}
\begin{figure}[t]
\begin{center}
\includegraphics[scale=0.3]{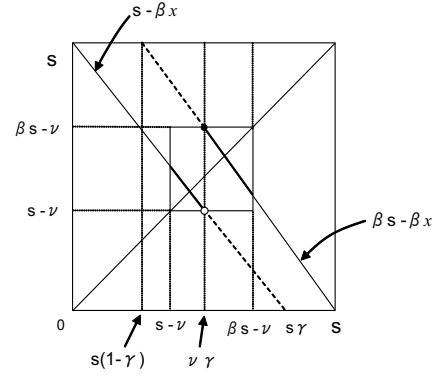} 
\end{center}
\caption{The map of the negative $\beta$-expansion:$R_{\beta, \nu, s}(x)$. 
Its invariant subinterval is a function of $\nu$ (see 
Figs. \ref{nbem123}(a),(b),(c) and \ref{Invariant subinterval}(b)).}
\label{NBM}
\end{figure}
\begin{figure}[htbp]
\begin{center}
\includegraphics[scale=0.30]{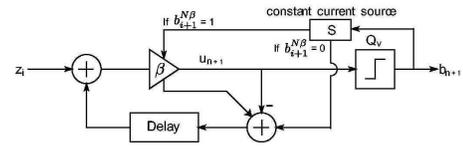}
\end{center}
\caption{The negative $\beta$-encoder.}
\label{NBE}
\end{figure}
This section introduces a radix expansion of a real number 
in a negative real base, called a \textit{negative $\beta$-expansion},
and discusses  differences between the quantisation errors of negative 
$\beta$-expansions and those of (scale-adjusted) $\beta$-expansions 
as well as their MSE. 
Such a negative radix expansion is  unusual, and somewhat intricate. 
First, consider a (scale-adjusted) negative $\beta$-expansion as a map 
$R_{\beta, \nu, s}(x):[0,s) \rightarrow [0,s),\,s>0$, defined by
\begin{equation}
\hspace*{-4mm}
R_{\beta, \nu, s}(x) =  \left\{
\begin{array}{ll}
s-\beta x,&\hspace*{-2mm}
 x \in [0, \gamma \nu), \\
\beta s - \beta x,&\hspace*{-2mm} x \in [\gamma \nu, s), \\
\end{array}
\right.
\nu \in [s(\beta - 1), s]
\end{equation}
as shown in Fig. \ref{NBM}. Such a negative $\beta$-expansion defines 
a new A/D converter as shown in Fig.\ref{NBE}, called 
a \textit{negative $\beta$-encoder} which facilitates the implementation 
of stable analog circuits~\footnote{Personal communication with 
Prof. Yoshihiko Horio.} 
and  improves  the quantisation MSE both
in the greedy case $\nu=s(\beta-1)$ and in the lazy case $\nu=s$ 
as follows. 
Let $b_{i,R^i_{\beta, \nu_1^i, s}}(x)$ be the associated bit 
sequence for the threshold sequence $\nu_1^L$, defined as 
\begin{equation}
b_{i,R^i_{\beta, \nu_1^i, s}}(x) = 
\left\{
\begin{array}{ll}
0,       & R_{\beta, \nu_1^{i-1}, s}^{i-1}(x) \in [0, \gamma \nu_i), \\
1, & R_{\beta, \nu_1^{i-1}, s}^{i-1}(x) \in [\gamma \nu_i, s), 
\end{array}
\right.
\label{eq:negativebit}
\end{equation}
where $R^i_{\beta, \nu_1^i, s}(x)$ is defined recursively as 
\begin{eqnarray}
&&R^L_{\beta,\nu_1^L}(x)=R_{\beta,\nu_L}(R^{L-1}_{\beta,\nu_1^{L-1}}(x))
\nonumber\\
&&
=R_{\beta,\nu_L}(R_{\beta,\nu_{L-1}}(R^{L-2}_{\beta,\nu_1^{L-2}}
(\cdots(R_{\beta,\nu_1}(x))).
\end{eqnarray}
The  scale-adjusted negative $\beta$ map $R_{\beta, \nu, s}(x)$ also defines 
the flaky quantiser $Q^f_{[s(1-\gamma),s\gamma]}(\cdot)$.
Then $R_{\beta, \nu_1^L, s}^L(x)$ can be represented recursively as follows: 
\begin{eqnarray}
\hspace*{-8mm}
&&\hspace*{-2mm}
R_{\beta, \nu, s}^L(x)=
  sf_{L,R^L_{\beta,\nu_1^L,s}}
-\beta R_{\beta, \nu_1^{L-1}, s}^{L-1}(x) \nonumber \\
&=&\hspace*{-2mm} sf_{L,R^L_{\beta,\nu_1^L,s}}
-\beta sf_{L-1,R^{L-1}_{\beta,\nu_1^{L-1},s}}
\nonumber \\
\hspace*{-8mm}
&+&\hspace*{-2mm}\cdots + 
(-\beta)^{L-1} sf_{1,R_{\beta,\nu_1^1,s}}
 + (-\beta)^L x \nonumber \\
\hspace*{-8mm}
&=&\hspace*{-2mm} s\sum^L_{i=1}f_{i,R^i_{\beta,\nu_1^i,s}}
(-\beta)^{L-i} + (-\beta)^L x
\end{eqnarray}
which yields
\begin{equation}
x = (-\gamma)^L R^L_{\beta,\nu_1^L,s}(x) -  
s\sum^L_{i=1}f_{i,R^i_{\beta,\nu_1^i,s}}(-\gamma)^{i},
\label{RBX}
\end{equation}
where 
\begin{equation}
f_{i,R^i_{\beta,\nu_1^i,s}}=1+b_{i,R^i_{\beta,\nu_1^i,s}}(\beta-1).
\label{RBX}
\end{equation}
The relation $R^L_{\beta, \nu_1^L, s}(x) \in [0, s)$ defines 
its subinterval, 
\begin{eqnarray}
\hspace*{-5mm}
&&I_{2L+1,R^{2L+1}_{\beta,\nu_1^{2L+1},s}}(x)\nonumber\\
\hspace*{-12mm}
&=&\hspace*{-3mm}
s[-\gamma^{2L+1}-\sum^{2L+1}_{i=1}
f_{i,R^i_{\beta,\nu_1^i,s}}(-\gamma)^i, 
-\sum^{2L+1}_{i=1}f_{i,R^i_{\beta,\nu_1^i,s}}(-\gamma)^i)\nonumber\\
&&\\
\hspace*{-6mm}
&&I_{2L, R^{2L}_{\beta,\nu_1^{2L},s}}(x) \nonumber\\
\hspace*{-9mm}
&= &\hspace*{-3mm}
s[-\sum^{2L}_{i=1}f_{i,R^i_{\beta,\nu_1^i,s}}(-\gamma)^i, 
\gamma^{2L}
-\sum^{2L}_{i=1}f_{i,R^i_{\beta,\nu_1^i,s}}(-\gamma)^i)
\label{eq:rmapinterval}
\end{eqnarray}
which  enables us to obtain 
the following decoded value $\widehat{x}_{L,R_{\beta,\nu_1^L,s}}$:
\begin{equation}
\widehat{x}_{L,R^L_{\beta,\nu_1^L,s}} = s\{ (-\gamma)^L /2 - 
\sum^L_{i=1}f_{i,R^i_{\beta,\nu_1^i,s}}(-\gamma)^{i}\}.
\label{RBD}
\end{equation}
Then, its quantisation error $\epsilon_{L,R^L_{\beta,\nu_1^L,s}}(x)$ 
is bounded as
\begin{equation}
\epsilon_{L,R^L_{\beta,\nu_1^L,s}}(x) = 
|x -\widehat{x}_{L,R^L_{\beta,\nu_1^L,s}}|\le \dfrac{s \gamma^L}{2},
\end{equation}
which is the same as in the scale-adjusted $\beta$-expansion.

Such a successive approximation of $x$  by the cautious expansion using 
$\nu_1^L$ is described by the contraction process of the interval by 
the negative $\beta$-expansion as follows.
Let $I_{i, R^i_{\beta,\nu_1^i,s}}(x)=[l^R_i(\nu_1^i),r^R_i(\nu_1^i)),\,i\geq1$ 
be the $i\mbox{th}$ interval for the negative $\beta$-expansion map 
$R_{\beta,\nu_i,s}(x)$\,with the threshold sequence $\nu_1^i, 
(s(\beta-1)\leq\nu_j\leq s)$, recursively defined by
\begin{equation}
\begin{array}{lcr}
l^R_{2\ell+1}(\nu_1^{2\ell+1})&=&l^R_{2\ell}(\nu_1^{2\ell})-s\gamma^{2\ell+1}
(1-f_{2\ell+1,R^{2\ell+1}_{\beta,\nu_1^{2\ell+1},s}}),\\
r^R_{2\ell+1}(\nu_1^{2\ell+1})&=&r^R_{2\ell}(\nu_1^{2\ell})-s\gamma^{2\ell}
(1-f_{2\ell+1,R^{2\ell+1}_{\beta,\nu_1^{2\ell+1},s}}\gamma),\\
l^R_{2\ell}(\nu_1^{2\ell})&=&l^R_{2\ell-1}(\nu_1^{2\ell-1})+s\gamma^{2\ell-1}
(1-f_{2\ell,R^{2\ell}_{\beta,\nu_1^{2\ell},s}}\gamma),\\
r^R_{2\ell}(\nu_1^{2\ell})&=&r^R_{2\ell-1}(\nu_1^{2\ell-1})+s\gamma^{2\ell}
(1-f_{2\ell,R^{2\ell}_{\beta,\nu_1^{2\ell},s}}).\\
\end{array}
\label{eq:recursionfomulanegative}
\end{equation}
Thus, we have the following lemma.

{\bf Lemma $4$}\,: 
{\it If $x\in I_{i,R^i_{\beta,\nu_1^i,s}}(x)$, then $x\in 
I_{i+1,R^{i+1}_{\beta,\nu_1^{i+1},s}}(x),\,\, i\in\mathbb{Z}$.
}

{\itshape proof:}
It is obvious that $x\in I_{0,R_{\beta,\nu,s}}(x)=[0,s)$ because 
$x\in[0,1)\subset I_{0,R_{\beta,\nu}}(x)$.
Suppose that $x\in I_{2\ell,R^{2\ell}_{\beta,\nu_1^{2\ell},s}}(x)$. 
Then, 
\begin{eqnarray*}
\hspace*{-4mm}&&\hspace*{-2mm} x-l^R_{2\ell+1}(\nu_1^{2\ell+1})\\
\hspace*{-4mm}&=&\hspace*{-2mm}
\gamma^{2\ell}R^{2\ell}_{\beta,\nu_1^{2\ell},s}(x)
-s\displaystyle{\sum_{i=1}^{2\ell}f_{i,R^i_{\beta,\nu_1^{i},s}}(-\gamma)^{i}}
\\
\hspace*{-2mm}&-&\hspace*{-2mm}
l^R_{2\ell}(\nu_1^{2\ell})+s\gamma^{2\ell+1}
(1-f_{2\ell+1,R^{2\ell+1}_{\beta,\nu_1^{2\ell+1},s}})\\
\hspace*{-4mm}&=&\hspace*{-2mm}
\gamma^{2\ell}
\{R^{2\ell}_{\beta,\nu_1^{2\ell},s}(x)
+s\gamma(1-f_{2\ell+1,R^{2\ell+1}_{\beta,\nu_1^{2\ell+1},s}})\},
\end{eqnarray*}
which suggests that if 
$R^{2\ell}_{\beta,\nu_1^{2\ell},s}(x)<\gamma\nu_{2\ell+1},\mbox{{\it i.e.,}}$ 
$f_{2\ell+1,R^{2\ell+1}_{\beta,\nu_1^{2\ell+1},s}}=1$, 
then $x-l^R_{2\ell+1}(\nu_1^{2\ell+1})=
\gamma^{2\ell}R^{2\ell}_{\beta,\nu_1^{2\ell},s}(x)\geq0$. 
Otherwise 
$\mbox{{\it i.e.,}}\, f_{2\ell+1,R^{2\ell+1}_{\beta,\nu_1^{2\ell+1},s}}=\beta$, $x-l^R_{2\ell+1}(\nu_1^{2\ell+1})\geq\gamma^{2\ell+1}
\{\nu_{2\ell+1}-s(\beta-1)\}
\geq0$ 
because $\nu_{2\ell+1}\geq s(\beta-1)$. 
On the other hand, 
\begin{eqnarray*}
\hspace*{-6mm}&&\hspace*{-3mm} r^R_{2\ell+1}(\nu_1^{2\ell+1})-x\\
\hspace*{-6mm}&=&\hspace*{-3mm}
r^R_{2\ell}(\nu_1^{2\ell})-s\gamma^{2\ell}
(1-f_{2\ell+1,R^{2\ell+1}_{\beta,\nu_1^{2\ell+1},s}}\gamma)
\\
\hspace*{-6mm}&-&\hspace*{-3mm}
\{\gamma^{2\ell}R^{2\ell}_{\beta,\nu_1^{2\ell},s}(x)
-s\displaystyle{\sum_{i=1}^{2\ell}
f_{i,R^{i}_{\beta,\nu_1^{i},s}}(-\gamma)^{i}}
\}\\
\hspace*{-6mm}&=&\hspace*{-3mm}
s\gamma^{2\ell+1}f_{2\ell+1,R^{2\ell+1}_{\beta,\nu_1^{2\ell+1},s}}
-\gamma^{2\ell}R^{2\ell}_{\beta,\nu_1^{2\ell},s}(x)
\\\hspace*{-6mm}&=&\hspace*{-2mm}
\gamma^{2\ell}\{s\gamma f_{2\ell+1,R^{2\ell+1}_{\beta,\nu_1^{2\ell+1},s}}
-R^{2\ell}_{\beta,\nu_1^{2\ell},s}(x)\},
\end{eqnarray*}
which suggests that if 
$R^{2\ell}_{\beta,\nu_1^{2\ell},s}(x)<\gamma\nu_{2\ell+1}$,
$\mbox{{\it i.e.,}}$ $f_{2\ell+1,R^{2\ell+1}_{\beta,\nu_1^{2\ell+1},s}}=1$, 
then 
$r^R_{2\ell+1}(\nu_1^{2\ell+1})-x
\geq\gamma^{2\ell+1}(s-\nu_{2\ell+1})\geq0$ 
because $\nu_{2\ell+1}<s$.
Otherwise\, $\mbox{{\it i.e.,}}\, 
f_{2\ell+1,R_{\beta,\nu_1^{2\ell+1},s}}=\beta$, 
$r^R_{2\ell1}(\nu_1^{2\ell+1})-x\geq\gamma^{2\ell}
\{s-R^{2\ell}_{\beta,\nu_1^{2\ell},s}(x)\}\geq0$
 because $R^{2\ell}_{\beta,\nu_1^{2\ell},s}(x))\leq s$.
This implies that 
$x\in I_{2\ell+1,R^{2\ell+1}_{\beta,\nu_1^{2\ell+1},s}}(x).$\\
Similarly, suppose that 
$x\in I_{2\ell-1,R^{2\ell-1}_{\beta,\nu_1^{2\ell-1}}}(x)$.
Then, 
\begin{eqnarray*}
\hspace*{-4mm}&&\hspace*{-2mm} x-l^R_{2\ell}(\nu_1^{2\ell})\\
\hspace*{-4mm}&=&\hspace*{-2mm}
-\gamma^{2\ell-1}R^{2\ell-1}_{\beta,\nu_1^{2\ell-1},s}(x)
-s\displaystyle{\sum_{i=1}^{2\ell-1}
f_{i,R^{i}_{\beta,\nu_1^{i},s}}(-\gamma)^{i}}
\\\hspace*{-4mm}&+&\hspace*{-2mm}
s\displaystyle{\sum_{i=1}^{2\ell}
f_{i,R^{i}_{\beta,\nu_1^{i},s}}(-\gamma)^{i}}\\
\hspace*{-10mm}&=&\hspace*{-2mm}
\gamma^{2\ell-1}
\{s\gamma f_{2\ell,R_{\beta,\nu_1^{2\ell},s}^{2\ell}}
-R^{2\ell-1}_{\beta,\nu_1^{2\ell-1},s}(x)\},
\end{eqnarray*}
which suggests that if 
$R^{2\ell-1}_{\beta,\nu_1^{2\ell-1}}(x)<
\gamma\nu_{2\ell},\,\mbox{{\it i.e.,}}\, 
f_{2\ell,R_{\beta,\nu_1^{2\ell},s}^{2\ell}}=1$, then 
$x-l^R_{2\ell}(\nu_1^{2\ell})=
\gamma^{2\ell}(s-\nu_{2\ell})
\geq0$ because 
$\nu_{2\ell}\leq s$.
Otherwise\, $\mbox{{\it i.e.,}} 
f_{2\ell,R_{\beta,\nu_1^{2\ell},s}^{2\ell}}=\beta$, 
$x-l^R_{2\ell}(\nu_1^{2\ell})=\gamma^{2\ell-1}
\{s-R^{2\ell-1}_{\beta,\nu_1^{2\ell-1},s}(x)\}\geq0$
 because 
$R^{2\ell-1}_{\beta,\nu_1^{2\ell-1},s}(x)\leq s$.
On the other hand, 
\begin{eqnarray*}
\hspace*{-4mm}&&\hspace*{-2mm} r^R_{2\ell}(\nu_1^{2\ell})-x\\
\hspace*{-4mm}&=&\hspace*{-2mm}
r^R_{2\ell-1}(\nu_1^{2\ell-1})+s\gamma^{2\ell}
(1-f_{2\ell,R^{2\ell}_{\beta,\nu_1^{2\ell},s}})\\
\hspace*{-2mm}&+&\hspace*{-2mm}
\gamma^{2\ell-1}R^{2\ell-1}_{\beta,\nu_1^{2\ell-1},s}(x)
+s\displaystyle{\sum_{i=1}^{2\ell-1}
f_{i,R^{i}_{\beta,\nu_1^{i},s}}(-\gamma)^{i}}\\
\hspace*{-2mm}&=&\hspace*{-2mm}
\gamma^{2\ell-1}\{s\gamma
(1-f_{2\ell,R^{2\ell}_{\beta,\nu_1^{2\ell},s}})
+R^{2\ell-1}_{\beta,\nu_1^{2\ell-1},s}(x)\},
\end{eqnarray*}
which implies that 
if $R^{2\ell-1}_{\beta,\nu_1^{2\ell-1},s}(x)<\gamma\nu_{2\ell}, 
\mbox{{\it i.e.,}} 
f_{2\ell,R^{2\ell}_{\beta,\nu_1^{2\ell},s}}=1$, then 
$r_{2\ell}^R(\nu_1^{2\ell},s)-x=
\gamma^{2\ell-1}R^{2\ell-1}_{\beta,\nu_1^{2\ell-1},s}(x)\geq0$. 
Otherwise,\, $\mbox{{\it i.e.,}} 
f_{2\ell,R^{2\ell}_{\beta,\nu_1^{2\ell},s}}=\beta, 
r^R_{2\ell}(\nu_1^{2\ell})-x\geq
\gamma^{2\ell}\{-s(\beta-1)+\nu_{2\ell}\}\geq0$
 because $s(\beta-1)\leq\nu_{2\ell}$.
This implies that 
$x\in I_{2\ell,R^{2\ell}_{\beta,\nu_1^{2\ell}},s}(x).$
This completes the proof.\hfill $\Box$

Similarly, using the idea of Daubechies {\it et al.} 
and  the negative $\beta$-expansion sequences $b_{i,R^i_{\beta,\nu_1^i,s}}$ 
for $x$ and $c_{i,R^i_{\beta,\nu_1^i,s}}$ for $y = 1 - x$ $(i=1,2,\cdots,L)$,
we can get the characteristic equation of $\beta$ in a 
negative $\beta$-encoder as follows:
\begin{eqnarray}
\hspace*{-6mm}&&\hspace*{-3mm}P_{L,R^L_{\beta,\nu_1^L,s}}(\gamma)
\nonumber\\ %
\hspace*{-6mm}&=&\hspace*{-3mm}
s\{(-\gamma)^L-\hspace*{-1mm}\sum^{L}_{i=1} 
(2+\{b_{i,R^i_{\beta,\nu_1^{i},s}}+
c_{i,R^i_{\beta,\nu_1^i,s}}\}(\beta-1))(-\gamma)^i\}-1\nonumber \\
\hspace*{-6mm} &=& \hspace*{-3mm}
s\{[d_{L,R^L_{\beta,\nu_1^{L},s}}-1](-\gamma)^L\nonumber\\
\hspace*{-6mm}&-& \hspace*{-3mm}
\hspace*{-1mm}\sum^{L-1}_{i=1}
[2-d_{i,R^i_{\beta,\nu_1^{i},s}}
+d_{i+1,R^{i+1}_{\beta,\nu_1^{i+1},s}}](-\gamma)^i\}-1,
\label{eq:negativebetacharacteristic equation}
\end{eqnarray}
where $d_{i,R^i_{\beta,\nu_1^{i},s}}=
b_{i,R^i_{\beta,\nu_1^{i},s}}+c_{i,R^i_{\beta,\nu_1^i,s}}$.
However, it is hard to guarantee the uniqueness of the root of 
Eq.(\ref{eq:negativebetacharacteristic equation}). 

\begin{figure}[htbp]
\begin{center}
\hspace*{-12mm}
\includegraphics[scale=0.35]{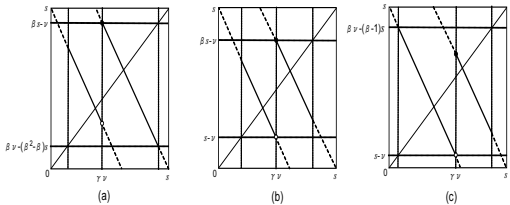} 
\end{center}
\caption{Negative $\beta$-expansion map $R_{\beta,\nu,s}(x)$ with its invariant subinterval when 
(a) $(\beta - 1) s \le \nu < \dfrac{\beta^2 -\beta + 1}{\beta + 1} s$,
(b) $\dfrac{\beta^2 -\beta + 1}{\beta + 1} s \le \nu < \dfrac{2 \beta - 1}{\beta + 1} s$,
and 
(c) $\dfrac{2 \beta - 1}{\beta + 1}  s \le \nu \le s $.}
\label{nbem123}
\end{figure}

The invariant subinterval of the map $R_{\beta, \nu, s}(x)$ 
consists of 2 line segments as shown in Fig.\ref{NBM}. 
The line segment in the graph 
of $R_{\beta, \nu, s}(x)$ with the full range is called 
a {\it full line segment}.
\footnote{In the $\beta$-expansion, the greedy-expansion map (or the lazy-expansion map), restricted to its invariant subinterval has a left (or right) full 
line segment as shown in Fig.\ref{greedylazymap} (a) (or (b)) but the cautious-expansion map, restricted to its invariant subinterval has no full line segment as shown in Fig.\ref{cautiousmap} (a).}  
There are three possible cases as follows:
\begin{enumerate}
\item   The right segment is  the full line segment 
(as shown in Fig. \ref{nbem123}(a)) whose invariant subinterval is given by 
\begin{equation}
[\beta \nu - (\beta^2 - \beta) s, \beta s - \nu) 
\end{equation}
when
$(\beta - 1)  s \le \nu < \dfrac{\beta^2 -\beta + 1}{\beta + 1} s $
since
\begin{eqnarray}
&& \hspace*{-14mm}
\beta s - \nu - \{ \beta \nu - (\beta^2 -  \beta) s \}
> \beta s - \nu - \{ s - \nu \} 
\nonumber \\
\hspace*{-10mm}& \Leftrightarrow &
(\beta^2 -\beta + 1)s - (\beta + 1) \nu > 0
\nonumber \\
\hspace*{-10mm}& \Leftrightarrow &
\nu < \dfrac{\beta^2 - \beta + 1}{\beta + 1}s.
\nonumber
\end{eqnarray}
\item No full line segment (as shown in Fig. \ref{nbem123}(b)) 
whose invariant subinterval is given by 
\begin{equation}
[s - \nu, \beta s - \nu), 
\end{equation}
when $\dfrac{\beta^2 -\beta + 1}{\beta + 1} s \le \nu < 
\dfrac{2 \beta - 1}{\beta + 1} s$ 
since 
\begin{eqnarray}
&& \hspace*{-10mm}
\beta s-\nu - (s-\nu)=(\beta - 1)s 
> \beta\nu -(\beta-1)s-(s - \nu) 
\nonumber \\
\hspace*{-10mm}& \Leftrightarrow &
(2\beta -1)s >(\beta +1)\nu.
\nonumber \\
& \Leftrightarrow &
\nu<\dfrac{2 \beta - 1}{\beta + 1}s.
\nonumber
\end{eqnarray}
\item The left segment is the full line segment 
(as shown in Fig. \ref{nbem123}(c)) whose invariant subinterval is given by 
\begin{equation}
[s - \nu, \beta \nu - s(\beta - 1)),
\end{equation}
when
$\dfrac{2 \beta - 1}{\beta + 1}  s \le \nu \le s $ 
since
\begin{eqnarray}
&& \hspace*{-10mm}
\beta \nu - (\beta - 1)s  - \{ s - \nu \} 
\ge \beta s - \nu  - ( s - \nu )
\nonumber \\
\hspace*{-10mm}& \Leftrightarrow &
(\beta + 1) \nu - (2 \beta -1)s \ge 0
\nonumber \\
& \Leftrightarrow &
\dfrac{2 \beta - 1}{\beta + 1}s \le \nu.
\nonumber
\end{eqnarray}
\end{enumerate}
%
\begin{figure}[htbp]
\begin{center}
\hspace*{-8mm}\includegraphics[scale=0.45]{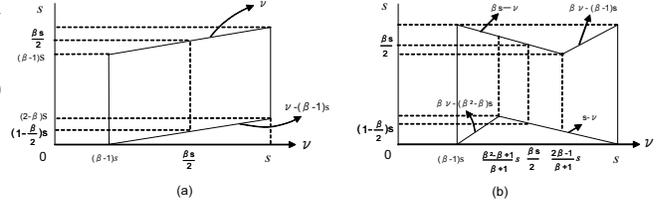} 
\end{center}
\caption{Invariant subinterval, a function of $\nu$, in (a) an ordinary 
$\beta$-expansion (see Fig.\ref{greedylazymap}(a),(b)) and (b) a negative $\beta$-expansion 
(see Fig.\ref{nbem123}(a),(b),(c)) .}
\label{Invariant subinterval}
\end{figure}
\begin{figure}[htbp]
\begin{center}
\includegraphics[scale=0.45]{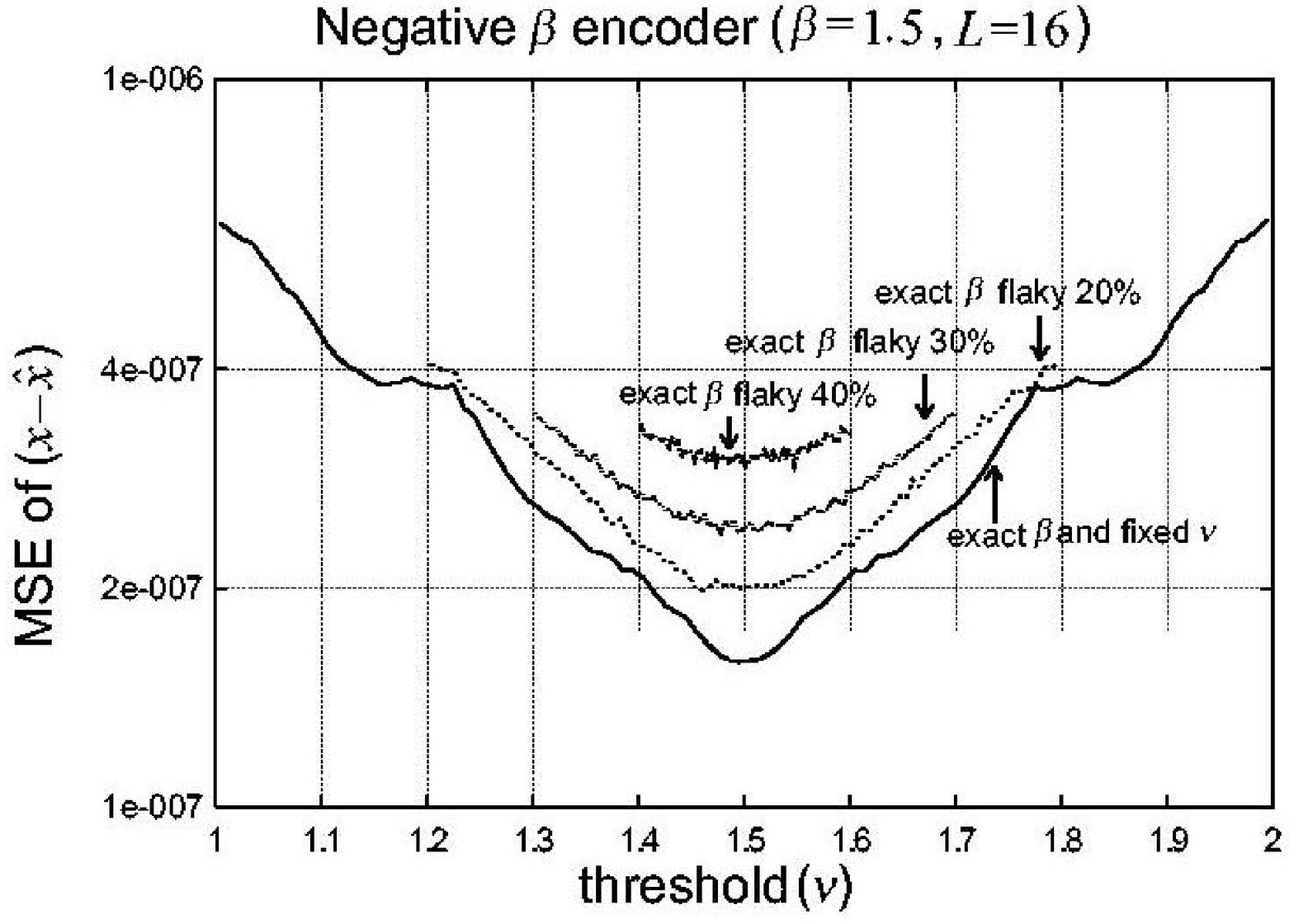} 
\end{center}
\caption{The $\mbox{MSE}(\widehat{x}_{L,R^L_{\beta,\nu_1^L,s}}(\gamma,1))$ 
using the exact $\beta$ of the negative $\beta$-encoder with 
 $\beta =1.5$, $L=16$, and $s=(\beta-1)^{-1}$ for 
fluctuating $\nu_i,\,1\leq i\leq L$\,
with its fluctuations $0,20,30,40\%$.}
\label{Varxp_L=1fluc}
\end{figure}
\begin{figure}[htbp]
\begin{center}
\includegraphics[scale=0.45]{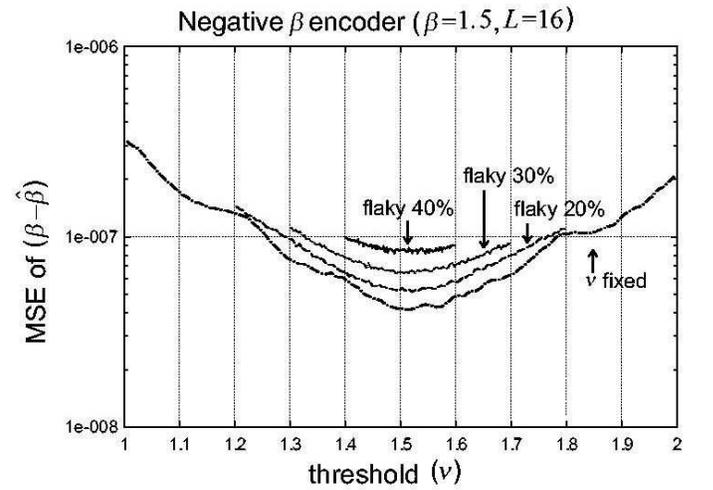} 
\end{center}
\caption{The MSE of the estimated $\beta$, 
{\it i.e.,} 
$\widehat{\beta}_{L,R^L_{\beta,\nu_1^L},s}(\gamma,1)$  
of the negative $\beta$-encoder with $\beta =1.5$, $L=16$ and 
$s=(\beta-1)^{-1}$ for fluctuating $\nu_i,\,1\leq i\leq L$\,
with its fluctuations $0,20,30,40\%$.
}
\label{Varbeta1neg}
\end{figure}
\begin{figure}[htbp]
\begin{center}
\includegraphics[scale=0.45]{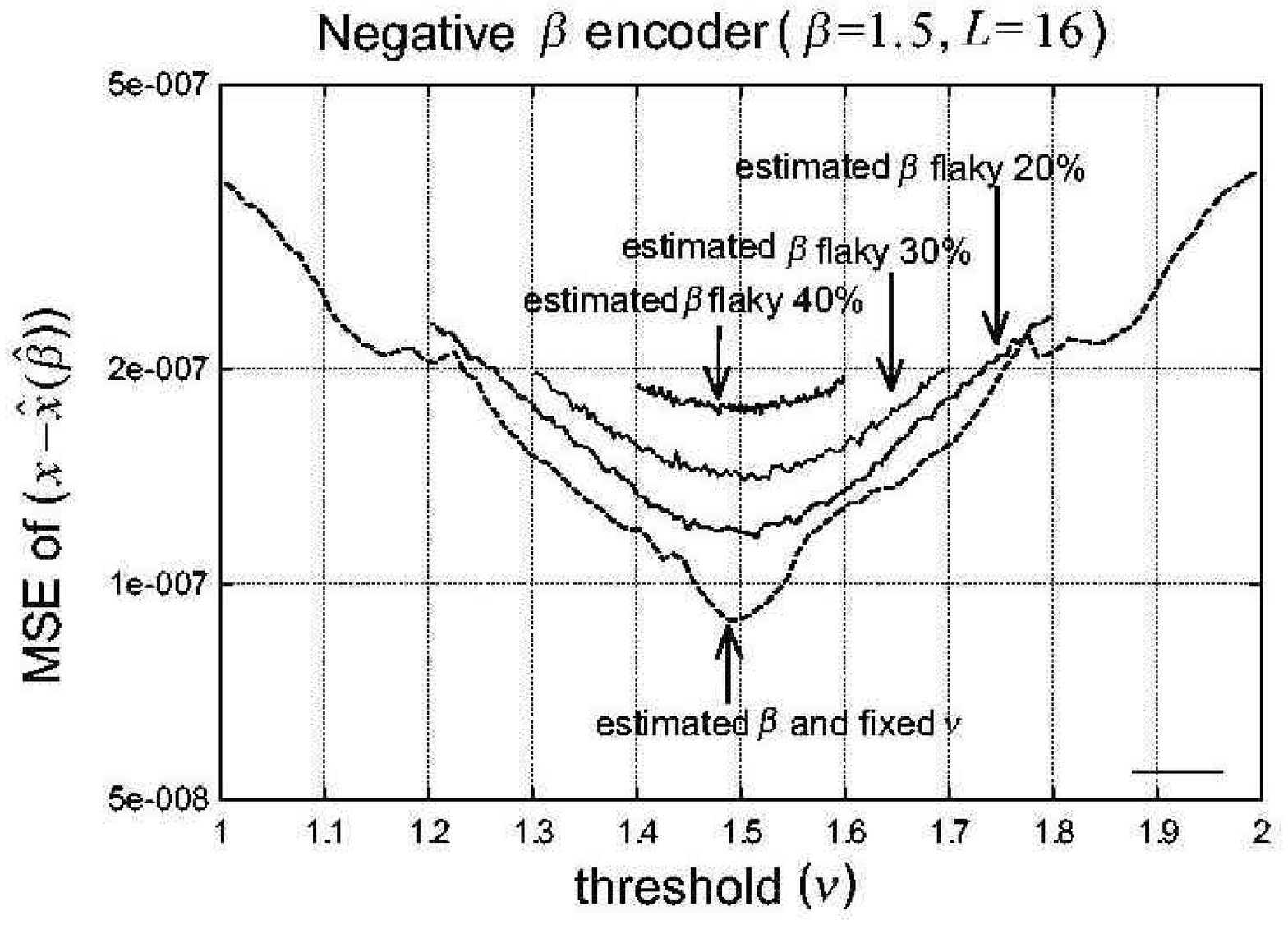} 
\end{center}
\caption{The $\mbox{MSE}(\widehat{x}_{L,R^L_{\beta,\nu_1^L,s}}(\widehat{\gamma},1))$  using the estimated $\widehat{\beta}$
 of the negative $\beta$-encoder with $\beta =1.5$, $L=16$, and 
$s=(\beta-1)^{-1}$ for fluctuating $\nu_i,\,1\leq i\leq L$\,
with its fluctuations $0,20,30,40\%$.}
\label{Varbeta1negfluct}
\end{figure}
\begin{figure}[htbp]
\begin{center}
\includegraphics[scale=0.45]{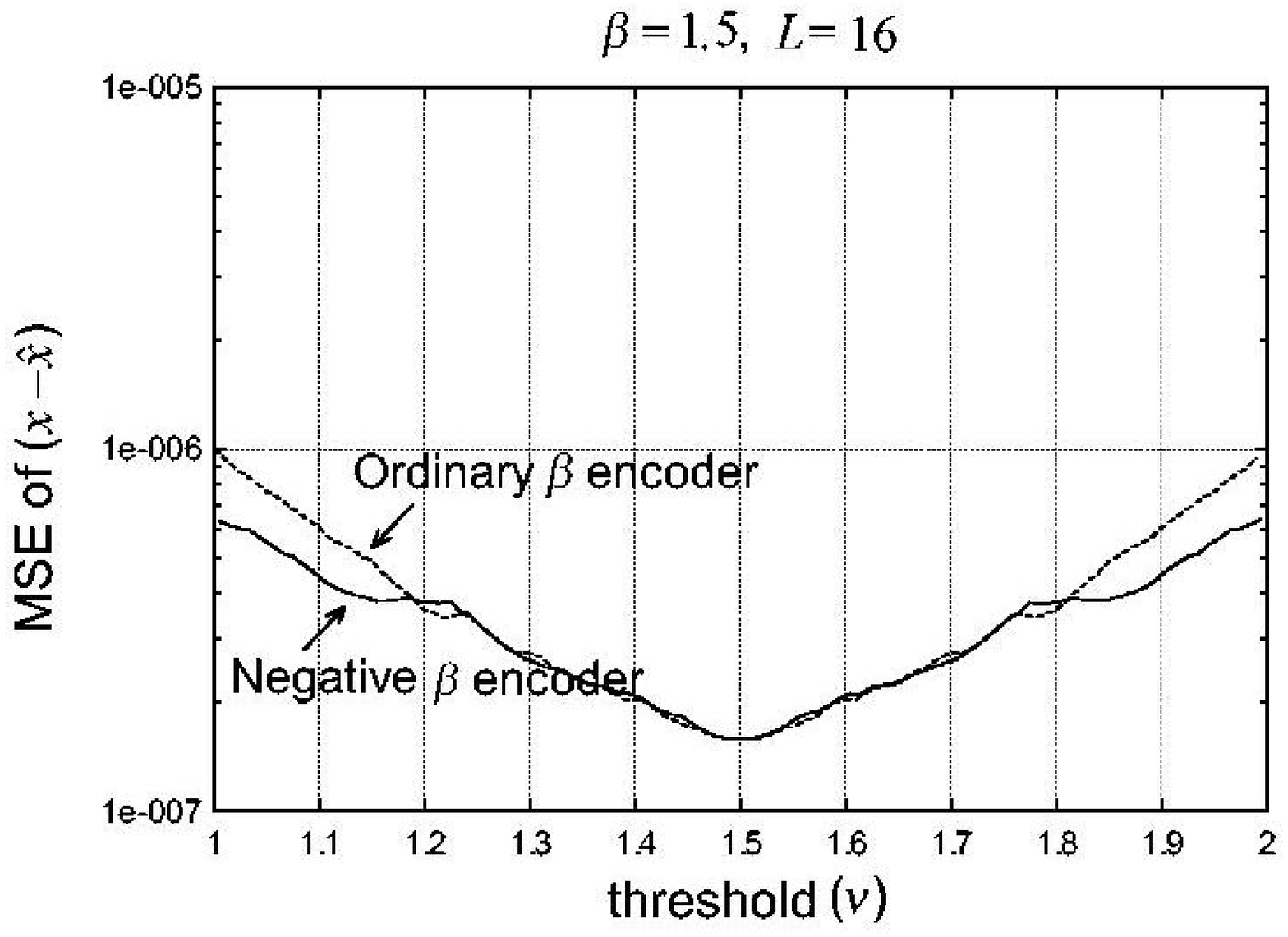} 
\end{center}
\caption
{The $\mbox{MSE}(\widehat{x}_{L,C^L_{\beta,\nu^*}}(\gamma,1))$ (or 
$\mbox{MSE}(\widehat{x}_{L,R^L_{\beta,\nu^*,s}}(\gamma,1))$ using the exact 
$\beta$  of the $\beta$-encoder (or the negative $\beta$-encoder) with 
 $\beta =1.5$, $L=16$, and \,$s=(\beta-1)^{-1}$ 
for fixed $\nu^*\in[1,(\beta-1)^{-1}]$.}
\label{Varxestimatebetaneg}
\end{figure}
\begin{figure}[htbp]
\begin{center}
\includegraphics[scale=0.45]{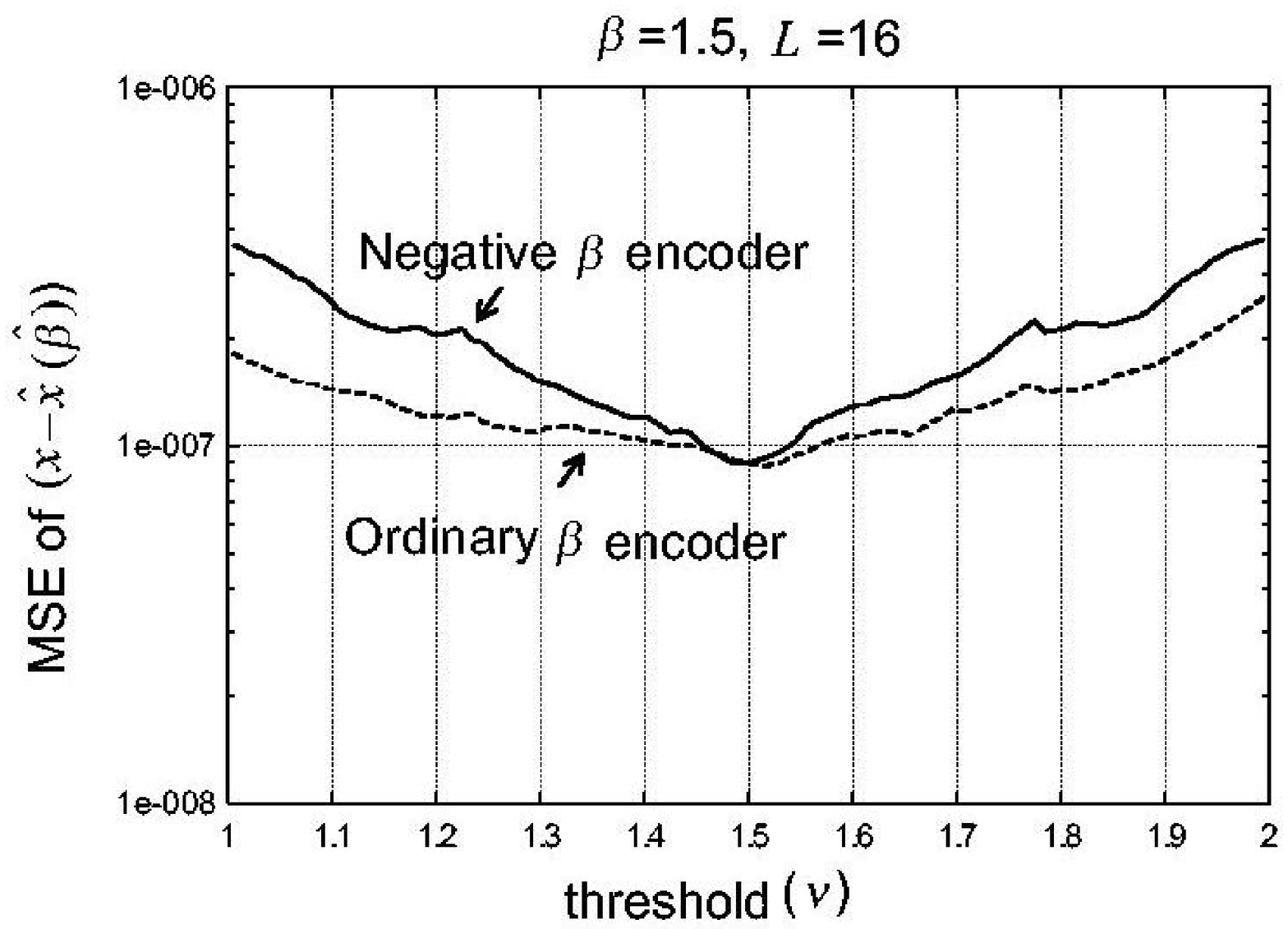} 
\end{center}
\caption
{The $\mbox{MSE}(\widehat{x}_{L,C^L_{\beta,\nu^*}}(\widehat{\gamma},1))$ 
(or $\mbox{MSE}(\widehat{x}_{L,R^L_{\beta,\nu^*,s}}(\widehat{\gamma},1))$\,
using the estimated $\widehat{\beta}$ 
 of the $\beta$-encoder (or the negative $\beta$-encoder) with 
$\beta =1.5$, $L=16$, and $s=(\beta-1)^{-1}$
for fixed $\nu^*\in[1,(\beta-1)^{-1}]$.}
\label{Varxestimatebetaneg1}
\end{figure}
\begin{figure}[htbp]
\begin{center}
\includegraphics[scale=0.45]{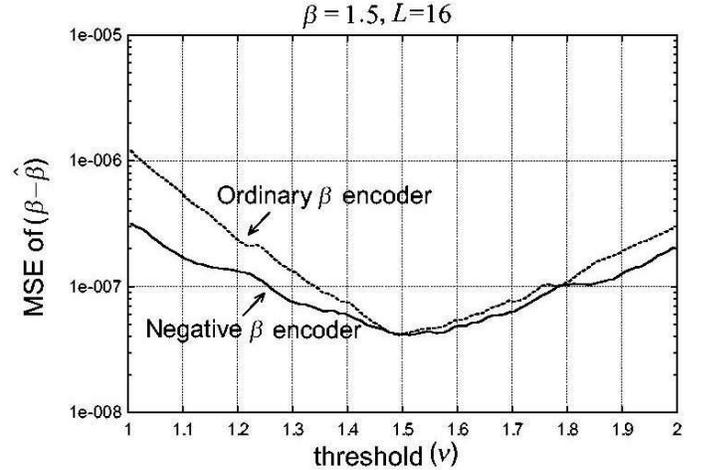} 
\end{center}
\caption{The MSE of the estimated $\beta$, {\it i.e.,} 
$\widehat{\beta}_{L,C^L_{\beta,\nu^*}}(\gamma,1)$  
( or $\widehat{\beta}_{L,R^L_{\beta,\nu^*}}(\gamma,1)$) 
of the $\beta$-encoder (or the negative $\beta$-encoder) 
with a fixed $\nu^*\in [1,(\beta-1)^{-1}]$ for $\beta =1.5$, $L=16$,
and $s=(\beta-1)^{-1}$.
}
\label{Varbeta1neg1}
\end{figure}
Since the quantisation error in the $\beta$-expansion is 
bounded as
\begin{equation}
\epsilon_{L, S^L_{\beta, \nu_1^L, s}}(x) = 
|x - \widehat{x}_{L,S^L_{\beta, \nu_1^L, s}}(x)|=
\gamma^L|\dfrac{s}{2} - S^L_{\beta, \nu_1^L, s}(x)|,
\end{equation}
the  MSE  of the quantisation decreases
if  $S_{\beta, \nu, s}(x)$ frequently takes on values  
in the middle portion of the interval $[0,s)$.
It is natural to assert that the quantisation threshold $\nu$ 
(with the inevitable errors) should be designed to be nearly equal to 
\begin{equation}
\nu=\beta s/2
\end{equation}
 so as to reduce the quantisation MSE (see Figs.~
\ref{Varxp_L=012},~\ref{Varxflakyp_L1estbeta},~\ref{Varbeta1},~\ref{Varxestimatebeta},~\ref{Varxfp_L2},~\ref{Varbeta2}). \\
The invariant subinterval in the $\beta$-expansion 
given as $[\nu - s(\beta - 1), \nu)$ as a function of $\nu$ is 
illustrated in Figure \ref{Invariant subinterval}(a), 
where  the $\beta$-expansion is  here called the {\it ordinary} 
$\beta$-expansion in order to discriminate between the $\beta$-expansion 
and the negative $\beta$-expansion. 
Hence the MSE becomes lower
 when $\nu = \dfrac{\beta s}{2}$ 
 since the invariant subinterval is given by $[(1-\dfrac{\beta}{2})s, 
\dfrac{\beta s}{2})$.
Meanwhile, the MSE in the greedy  expansion ($\nu = s(\beta - 1)$) 
(or the lazy expansion ($\nu = s$)) increases 
 because  the invariant subinterval is given as 
$[0, (\beta - 1)s)$ (or $[s(2 - \beta), s)$), 
which is skewed towards the left (or the right) 
 portion of the initial interval 
$[0,s)$ as shown in Fig.\ref{Invariant subinterval}(a).

On the other hand, the invariant subinterval in a  negative $\beta$-expansion as a function 
of $\nu$ is illustrated in Figure \ref{Invariant subinterval}(b). 
In particular, both of the greedy expansion  
and  the lazy expansion  have as 
their invariant subintervals, given as $[0,s)$, which is 
the same as the initial subinterval.
Therefore, the quantisation MSE automatically
becomes lower compared to that of a $\beta$-expansion,
while the invariant subinterval in a negative $\beta$-expansion 
with $\nu =\beta s/2$ is given by $[s(1-\beta/2),\beta s/2]$, 
which is the same as in a $\beta$-expansion 
with $\nu = \dfrac{\beta s}{2}$, so 
the MSE is comparable. 
Figures \ref{Varxp_L=1fluc} and \ref{Varbeta1neg}
show the $\mbox{MSE}(\widehat{x}_{L,R^L_{\beta,\nu_1^L,s}}(\gamma,1))$
using the exact $\beta$ of the negative $\beta$-encoder 
and the $\mbox{MSE}(\widehat{\beta}_{L,R^L_{\beta,\nu_1^L,s}}(\widehat{x},1))$, respectively, with $\beta =1.5$, $L=16$,\,$s=(\beta-1)^{-1}$ for 
fluctuating $\nu_i,\,1\leq i\leq L$\,
with their several fluctuation bounds. 
 Figure \ref{Varbeta1negfluct}  shows the 
$\mbox{MSE}(\widehat{x}_{L,R^L_{\beta,\nu_1^L,s}}(\widehat{\gamma},1))$ 
using the estimated $\widehat{\beta}$ of the negative $\beta$-encoder. 
Comparing Fig.\ref{Varxp_L=1fluc} with Fig.\ref{Varbeta1negfluct} shows that 
$\widehat{x}_{L,R^L_{\beta,\nu_1^L,s}}(
\widehat{\gamma},1)$ gives a better approximation to $x$ 
than $\widehat{x}_{L,R^L_{\beta,\nu_1^L,s}}(\gamma,1)$. 
 Figure \ref{Varxestimatebetaneg} shows the 
$\mbox{MSE}(\widehat{x}_{L,C^L_{\beta,\nu_1^L}}(\gamma,1))$  
(or $\mbox{MSE}(\widehat{x}_{L,R^L_{\beta,\nu_1^L,s}}(\gamma,1))$) 
of quantizations  using the exact $\beta$ of the $\beta$-encoder 
(or the negative $\beta$-encoder). 
Figure \ref{Varxestimatebetaneg1} shows the 
  $\mbox{MSE}(\widehat{x}_{L,C^L_{\beta,\nu_1^L}}(\widehat{\gamma},1))$ 
(or $\mbox{MSE}(\widehat{x}_{L,R^L_{\beta,\nu_1^L,s}}(\widehat{\gamma},1))$) using the estimated $\widehat{\beta}$ of the $\beta$-encoder 
(or the negative 
$\beta$-encoder) with $\,\beta =1.5$, $L=16$,\,$s=(\beta-1)^{-1}$ for fixed 
$\nu^*\in [1,(\beta-1)^{-1}]$.
 Note that the $\mbox{MSE}(\widehat{x}_{L,C^L_{\beta,\nu_1^L}}(\widehat{\gamma},1)))$ is smaller than the 
$\mbox{MSE}(\widehat{x}_{L,R^L_{\beta,\nu_1^L,s}}(\gamma,1)))$
even though the $\mbox{MSE}
(\widehat{\beta}_{L,R^L_{\beta,\nu_1^L,s}}(\widehat{x},1)))$ 
is smaller than the 
$\mbox{MSE}(\widehat{\beta}_{L,C^L_{\beta,\nu_1^L}}(\widehat{x},1)))$ 
as shown in Fig.\ref{Varbeta1neg1}.
%
\section{Markov chain of binary sequences generated by $\beta$-encoder}
\begin{figure}[bp]
\begin{center}
\includegraphics[scale=0.45]{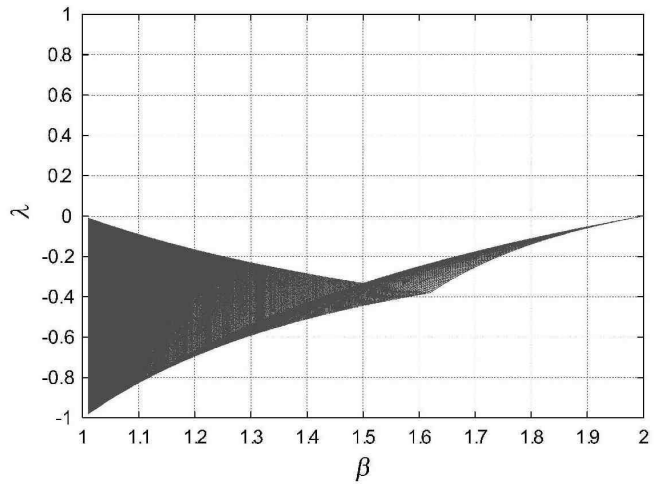}%
\end{center}
\caption{The second eigenvalue of the approximated transition probability 
matrix $P(\beta,\nu)$ as a function of $\beta$ and $\nu$.}
\label{eigenvalue1}
\end{figure}

As discussed above, 
the {\it localized} 
invariant subinterval makes the quantisation MSE in greedy/lazy-expansions 
worse than that of the cautious-expansion. 
 Regarding  a binary sequence generated by the greedy/lazy and cautious 
expansions as a Markov chain generating binary sequences, we show another 
clear distinction between them.
Furthermore, we observe that such a Markov chain explains the probabilistic behavior of the flaky quantisers, defined by the multivalued R\'enyi maps.\\
First, let us notice that there is a close relationship between information sources and a Markov chain 
with  a transition matrix of a finite dimension~\cite{ProcKohda}.
 The $\beta$-expansion maps, however, are not easy to characterize 
by transition matrices of size $2$ except for the original R\'enyi map with 
$\beta=(1+\sqrt{5})/2$.~\footnote{This value is a root of $\gamma=\beta-1$, 
which comes from the sufficient condition for the greedy map to be 
a two-state Markov map for the partition $\{[0,\gamma),[\gamma,1)\}$, i.e., 
the maps $[0,\gamma)\rightarrow[0,1)$ and $[\gamma,1)\rightarrow[0,\beta-1)$ 
satisfy the two-state Markov property (condition 3))
 of Definition $1$. 
} 
\\
Consider a special class of piecewise-linear Markov maps satisfying conditions 1)-4)  of Definition $1$, 
in which in addition, each $\tau|_{I_i}$ is required to be linear on $I_i$~\cite{Boyarsky}.
Such a map simply provides  a transition probability matrix.
~\footnote{Ulam~\cite{Ulam} posed the problem of the existence of an absolutely continuous invariant measure for the map, known as the {\it Ulam's conjecture}, and defined the transition probability matrix. 
See the Appendix for detail.} 
Kalman~\cite{Kalman} gave a {\it deterministic} procedure for embedding a 
Markov chain into the chaotic dynamics of the piecewise-linear-monotonic 
onto maps. 
Several attempts have been made to construct a dynamical system with 
an arbitrarily precribed Markov information source; in addition, 
by analogy with chaotic dynamics, 
arithmetic coding problems are also discussed~\cite{Barnsley,Kanaya}. 
The relationship between random number generation and interval algorithms 
has been discussed in~\cite{HanHoshi}. 
However, the few attempts or discussions in the following decades remind us 
that Kalman's  embedding procedure, as reviewed in the Appendix, 
is to be highly appreciated 
in the sense that Kalman addressed the question of whether 
irregular sequences observed in physical systems originated from 
{\it determinism} or not. \\
%
We turn now to the cautious expansion map $C_{\beta,\nu}(x)|_{[\nu-1,\nu)}$, 
consisting of two line segments $C_{\beta,\nu}(x)|_{[\nu-1,\gamma\nu)}$  and 
$C_{\beta,\nu}(x)|_{[\gamma\nu,\nu)}$. 
The inevitable fluctuations of $\beta$ and $\nu$ in an analog AD-conversion, 
however,  prevent the map $C_{\beta,\nu}(x)|_{[\nu-1,\nu)}$ from 
being a two-state 
Markov map for the partition $\{[\nu-1,\gamma\nu),[\gamma\nu,\nu)\}$, 
i.e., the maps 
$[\nu-1,\gamma\nu)\rightarrow[\beta(\nu-1),\nu)$ and 
$[\gamma\nu,\nu)\rightarrow[0,\beta\nu-1)$ 
 do not satisfy the $2$-state Markov property (see Fig.\ref{cautiousmap}(a)). 
This situation compels us to introduce an approximated transition matrix of size $2$ representing a $2$-state Markov chain~\cite{NDES2007} induced by the map $C_{\beta,\nu}(x)|_{[\nu-1,\nu)}$ as shown in Fig.\ref{cautiousmap}(a)
as follows.\\
A detailed observation of Fig.\ref{cautiousmap}(a) reveals that 
$C_{\beta,\nu}(x)<\gamma\nu,\,\mbox{when}\,x\in[\nu-1,\nu\gamma^2)\,\mbox{or}\, x\in[\nu\gamma,\nu\gamma^2+\gamma)$\, and\,
 $C_{\beta,\nu}(x)\geq\gamma\nu,\,\mbox{when}\,x\in[\nu\gamma^2,\nu\gamma)\,
\mbox{or}\, x\in[\nu\gamma^2+\gamma,\nu)$. 
Hence for 
$\displaystyle{\frac{\beta}{\beta^2-1}}\leq\nu<
\displaystyle{\frac{\beta^2}{\beta^2-1}}$,
this map $C_{\beta,\nu}(x)$ gives the following conditional 
probabilities: 
\begin{equation}
\begin{array}{lcr}
\hspace*{-4mm}\mbox{Pr}[X_{n+1}=0|X_n=0]=\displaystyle{\frac{\nu\gamma^2-(\nu-1)}{\nu\gamma-(\nu-1)}=1-\frac{U(\nu)}{\beta}},\\
\hspace*{-4mm}\mbox{Pr}[X_{n+1}=1|X_n=0]=\displaystyle{\frac{\nu\gamma-\nu\gamma^2}{\nu\gamma-(\nu-1)}=\frac{U(\nu)}{\beta}},\\
\hspace*{-4mm}\mbox{Pr}[X_{n+1}=0|X_n=1]=\displaystyle{\frac{\nu\gamma^2+\gamma-\nu\gamma}{\nu-\nu\gamma}=\frac{1}{\beta U(\nu)}},\\
\hspace*{-4mm}\mbox{Pr}[X_{n+1}=1|X_n=1]=\displaystyle{\frac{\nu-\nu\gamma^2-\gamma}{\nu-\nu\gamma}=1-\frac{1}{\beta U(\nu)}},
\end{array}
\end{equation}
where $U(\nu)=\displaystyle{\frac{S(\nu)}{T(\nu)}},S(\nu)=\nu(\beta-1)>0,\,$ and \,$T(\nu)=\nu-\beta(\nu-1)>0$. 
These propabilities define the transition matrix as follows: 
\begin{eqnarray}
\hspace*{-15mm}
&&P(\beta,\nu)\nonumber\\
\hspace*{-15mm}
&=&\hspace*{-3mm}\left\{\hspace*{-2mm}
\begin{array}{l}
\left(\hspace*{-2mm}
\begin{array}{ll}
1-\displaystyle{\frac{U(\nu)}{\beta}}
&
\displaystyle{\frac{U(\nu)}{\beta}}
\\
1&0
\end{array}
\hspace*{-2mm}\right),
\,\mbox{for}\,\nu<\frac{\beta}{\beta^2-1},\\
\left(\hspace*{-2mm}
\begin{array}{ll}
1-\displaystyle{\frac{U(\nu)}{\beta}}
&\displaystyle{\frac{U(\nu)}{\beta}}
\\
\displaystyle{\frac{1}{\beta U(\nu)}}
&1-\displaystyle{\frac{1}{\beta U(\nu)}}
\end{array}
\hspace*{-2mm}\right),
\,\mbox{for}\,\frac{\beta}{\beta^2-1}\leq\nu\leq\frac{\beta^2}{\beta^2-1},\\
\left(\hspace*{-2mm}
\begin{array}{ll}
0&1
\\
\displaystyle{\frac{1}{\beta U(\nu)}}
&
1-\displaystyle{\frac{1}{\beta U(\nu)}}
\end{array}
\hspace*{-2mm}\right),
\,\mbox{for}\,\nu>\frac{\beta^2}{\beta^2-1},
\end{array}
\hspace*{-8mm}\right.
\end{eqnarray}
whose stationary distribution is defined as
\begin{eqnarray}
\hspace*{-12mm}
&&(\mbox{Pr}[X=0],\,\mbox{Pr}[X=1])=(m_\infty,1-m_\infty)\nonumber\\
\hspace*{-6mm}
&&=\left\{\hspace*{-2mm}
\begin{array}{lcr} 
\displaystyle{\frac{1}{S(\nu)+\beta T(\nu)}}\left(\beta T(\nu),\,S(\nu)\right)
&\hspace*{-3mm}\mbox{for}\,\nu<\displaystyle{\frac{\beta}{\beta^2-1}},\\
\displaystyle{\frac{1}{S(\nu)^2+T(\nu)^2}}\left(T^2(\nu),\,S^2(\nu)\right)
&\hspace*{-3mm}\mbox{for}\,
\displaystyle{\frac{\beta}{\beta^2-1}}\leq\nu<\displaystyle{\frac{\beta^2}{\beta^2-1}},\\
\displaystyle{\frac{1}{T(\nu)+\beta S(\nu)}}\left(T(\nu),\,\beta S(\nu)\right)
&\hspace*{-3mm}\mbox{for}\,
\nu>\displaystyle{\frac{\beta^2}{\beta^2-1}},
\end{array}
\hspace*{-12mm}\right.\nonumber\\
\hspace*{-12mm}
&&\hspace*{-8mm}
\end{eqnarray}
and whose  second eigenvalue $\lambda(\beta,\nu)$, besides $1$, is given as
\begin{equation}
\hspace*{-4mm}
\lambda(\beta,\nu)\hspace*{-1mm}=\hspace*{-1mm}
\left\{\hspace*{-2mm}
\begin{array}{lcr} 
-\displaystyle{\frac{U(\nu)}{\beta }},&\hspace*{-2mm}\mbox{for}\,
\nu<\displaystyle{\frac{\beta}{\beta^2-1}},\\
1-\displaystyle{\frac{1}{\beta}}\left(U(\nu)+U^{-1}(\nu)
\right),&\hspace*{-2mm}\mbox{for}\,
\displaystyle{\frac{\beta}{\beta^2-1}}\leq\nu<\displaystyle{\frac{\beta^2}{\beta^2-1}},\\
-\displaystyle{\frac{1}{\beta U(\nu)}},&\hspace*{-2mm}\mbox{for}\,
\nu>\displaystyle{\frac{\beta^2}{\beta^2-1}}.
\end{array}
\hspace*{-8mm}\right.
\end{equation}
The second eigenvalue is bounded as $\lambda(\beta,\nu)\leq 1-
\displaystyle{\frac{2}{\beta}}\leq0$\\ 
if $\displaystyle{\frac{\beta}{\beta^2-1}}\leq\nu<\displaystyle{\frac{\beta^2}{\beta^2-1}}$ (or $\lambda(\beta,\nu)<0$ otherwise).\\
For the greedy case ($\nu=1$), we have 
$\lambda(\beta,1)=
1-\beta^{-1}(\beta-1+(\beta-1)^{-1})$ and 
$(\mbox{Pr}[X=0],\,\mbox{Pr}[X=1])
=\frac{1}{(\beta-1)^2+1}
\left(1,(\beta-1)^2\right)$; 
while for the lazy case ($\nu=(\beta-1)^{-1}$), we have 
$\lambda(\beta,(\beta-1)^{-1})=\lambda(\beta,1)$ and 
$(\mbox{Pr}[X=0],\,\mbox{Pr}[X=1])
=(1+\beta^2)^{-1}\left(\beta^2,\,1\right)$. 
 The second eigenvalues $\lambda(\beta,\nu)$ 
 illustrated in  Fig.\ref{eigenvalue1} show that 
for almost all $\beta$ and $\nu$, $P(\beta,\nu)$ 
has a negative eigenvalue of large magnitude except 
in the cases with $\beta=1$ and $\beta=2$. 
To confirm this fact, we introduce another method for estimating the non-unit 
eigenvalues of a two-state Markov chain as follows.

\begin{figure}[bp]
\begin{center}
\includegraphics[scale=0.45]{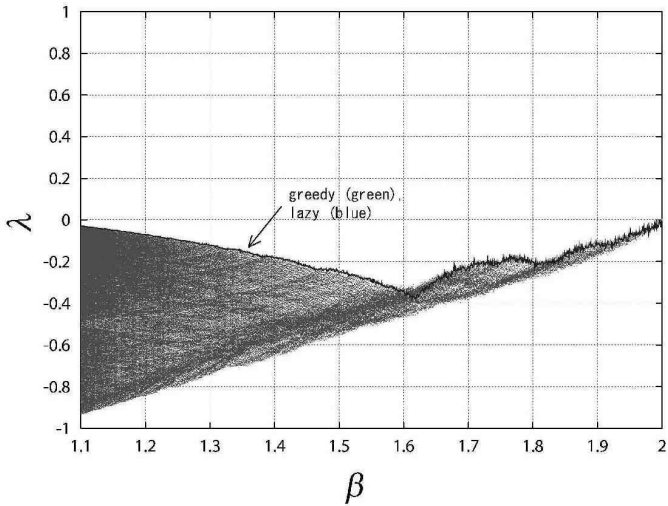}
\end{center}
\caption{The distribution of the second eigenvalue of the approximated 
transition probability $P(\{n_{ij}\})$ as a function of $\beta$ and $\nu$, 
for $N = 100,000$ and $x = \nu - \pi / 10$.}
\label{eigenvalue2}
\end{figure}

Let $b_1, b_2, \cdots , b_N$ be a binary sequence generated by 
the $\beta$-encoder. 
We regard $\{b_i\}$ as a two-state Markov chain with 
 transition matrix $P(\{n_{ij}\})$~\cite{NDES2007}, defined as  
\begin{equation}
P(\{n_{ij}\})=\left(
\begin{array}{*{2}{c}}
\frac{n_{00}}{n_{00} + n_{01}} & \frac{n_{01}}{n_{00} + n_{01}} \\
\frac{n_{10}}{n_{10} + n_{11}} & \frac{n_{11}}{n_{10} + n_{11}}
\end{array}
\right),
\end{equation}
where $n_{00}, n_{01}, n_{10}$ and $n_{11}$ are frequencies  
defined as
\begin{equation}
\hspace*{-6mm}
\begin{array}{lcl}
n_{00}=\sum^{N-1}_{i = 1} \overline{b_{i}}\cdot\overline{b_{i+1}},
&& \hspace*{-2mm}
n_{01}=\sum^{N-1}_{i = 1} \overline{b_{i}}\cdot b_{i+1}, \\
n_{10}=\sum^{N-1}_{i = 1} b_{i}\cdot\overline{b_{i+1}},
&&\hspace*{-2mm}
n_{11}=\sum^{N-1}_{i = 1} b_{i}\cdot b_{i+1}.
\end{array}
\end{equation}
Such a matrix $P(\{n_{ij}\})$ enables us 
to estimate the second eigenvalue $\tilde{\lambda}$ 
provided that $N$ is a sufficiently large number 
e.g. $N=100,000$.
The results illustrated in Fig.\ref{eigenvalue2} 
show that almost all the eigenvalues are negative 
and that the value $\tilde{\lambda}$ of the 
greedy (or the lazy) scheme is larger than that of the cautious scheme 
for almost all $\beta$ and $\nu$.
The negative non-unit eigenvalue of the transition probability matrix 
of size $2$, $P(\beta,\nu)$  plays an important role in designing spreading 
spectrum codes generated by a Markov chain with its negative eigenvalue 
in an asynchronous direct spread code multiple accesss system 
to improve the bit error performance~\cite{Bologna77,Bologna99,Kohda1}.
~\footnote{The reader interested in chaos-based spread-spectrum communication should see the review paper~\cite{ProcKohda} 
} 

\section{Conclusion}
The $\beta$-expansion has been shown to be characterized by the process 
of contraction of the subinterval containing a sample $x$. 
This has led to the following three results: 
(1) the new characteristic equation of the amplification factor $\beta$ 
 provides decoded values of $\beta$ and a sample $x$ with  high precisions;
(2) the negative $\beta$-encoder improves the quantisation MSE
in greedy/lazy schemes;
(3) if a binary sequence generated by the $\beta$-encoder is regarded as a two-state Markov chain, then the second eigenvalue of the Markov transition matrix is negative, and  the absolute value of the eigenvalue  is larger in the cautious scheme than in the greedy/lazy schmes, which  is relevant to  
the precision of the decoded values of both $\beta$ and $x$. 
However, it remains unknown why the value of x decoded using the estimated 
value $\widehat{\beta}$ gives a better approximation to $x$ than that using 
the exact value $\beta$. In addition, the relationship between 
the quantisation MSE and binary sequences, 
 approximated by Markov chains with the transition matrix having its 
negative eigenvalue has been omitted here 
because the MSE as a function of $\beta$ and $\nu$ is complicated 
 even for the transition probability matrix of size $2$, $P(\beta,\nu)$. 
In general, $\beta$-expansions need more sophisticated discussion using 
a Markov chain with a transition probability matrix of size more than $2$, 
which is an important problem for future research. 


%


\appendix[Kalman's  procedure of embedding a Markov chain into a nonlinear map]
Given a set of states $S=\{1,2,\cdots,N_s\}$
and a probability transition matrix $P=\{p_{ij}\}_{i,j=1}^{N_s}$,
satisfying $p_{ij}\geq0$ for all $i$, $j$;
$\sum_{j=1}^{N_s}p_{ij}=1$ for all $i$,
we define a sequence of random variables
$Z_0,Z_1,\ldots$ taking values in $S$.
If  $Z_0$ has an arbitrary distribution
\begin{equation}
\mbox{Pr}\{Z_{n+1}=s_k|Z_0=s_{i_0},\cdots,Z_n=s_{i_n}\}=p_{i_n,k},
\end{equation}
then the sequence of random variables $Z_0,Z_1,\ldots$
is called an $N_s$-state Markov chain.
%
Given a Markov chain $Z_0,Z_1,\ldots$ and a function $f$
whose domain is $S$ and whose range is an alphabet set
$\Gamma=\{\gamma_1,\cdots,\gamma_{N_a}\}$,
and assuming that the initial state $Z_0$ is chosen in accordance with
a stationary distribution ${\bf u}=(u_1,\cdots,u_{N_s})$,
then the stationary sequence $X_n=f(Z_n)$, $n=0,1,2,\ldots$
is said to be the Markov information source.
In this paper, for simplicity, we take $\Gamma=S$,
$N_a=N_s$, and $f$ to be the identity function.

Kalman gave a simple procedure for embedding a Markov chain
with transition matrix $P=\{p_{ij}\}_{i,j=1}^{N_s}$,
satisfying
\begin{equation}
0<p_{ij}<1 \quad \mbox{ for all } i,j
\label{eqn:positivepij}
\end{equation}
into an onto Piecewise-Linear Map (PLM)  map $\tau$ :
$ J=[0,1] \to J $ with
$N_s^2$ subintervals,
defined by
\begin{equation}
\omega_{n+1}=\tau(\omega_n),\quad n=0,1,2,\ldots,\quad \omega_n\in J
\end{equation}
as follows.~\footnote{ 
Readers interested in Ulam's conjecture, and Kalman's procedure as well as 
its revised version should see~\cite{ProcKohda}.}

First divide  the interval $J$ into $N_s$ subintervals such that
$J=\bigcup_{i=1}^{N_s}J_{i},$
where
\begin{equation}
\hspace*{-5mm}
J_{i}
=(d_{i-1},d_{i}],\quad
d_0=0<d_1<d_2<\cdots<d_{N_s}=1.
\label{eqn:Nainterval}
\end{equation}

Furthermore, divide the subintervals $J_{i}$ ($1\leq i\leq N_s$) into
$N_s$ subintervals such that
$J_{i}
=\bigcup_{j=1}^{N_s}J_{i,\,j},$
where
\begin{equation}
J_{i,\,j}
= \left\{
\begin{array}{l}
(d_{i,j-1},d_{i,j}]\quad
\mbox{ for }\,\tau(d_{i})=1,\\
\quad (d_{i,0}=d_{i-1},
\,\,d_{i,N_s}=d_{i}),
\vspace{1.5mm}\\
(d_{i,j},d_{i,j-1}]\quad
\mbox{ for }\,\tau(d_{i})=0,\\
\quad(d_{i,0}=d_{i},
\,\,d_{i,N_s}=d_{i-1}),
\end{array}
                     \right.
\label{eqn:interval}
\end{equation}
subject to the conditions of a Markov partition,
\begin{eqnarray}
\tau(d_{i})&\in& \{0,1\}, 
\quad\mbox{ for } 1\leq i \leq N_s,
\label{eqn:tauonto}\\
\tau(d_{i,j})&=&d_{j},
\quad\mbox{ for } 1\leq i,j \leq N_s,
\label{eqn:taudij}
\end{eqnarray}
and the condition of the transition probabilities
\begin{equation}
\frac{|J_{i,j}|}
{|J_i|}
=p_{ij},\quad\mbox{ for } 1\leq i,j \leq N_s.
\label{eqn:probij}
\end{equation}
Thus, the restrictions of Kalman's maps $\tau$ 
to the interval $J_{i,j}$, denoted by $\tau_{i,j}(\omega)$,
are of the form
\begin{equation}
\tau_{i,j}(\omega)
=
\left\{
\begin{array}{r}
\displaystyle{
\frac{|J_i|\omega
+(d_{i,j}d_{j-1}
-d_{i,j-1}d_{j})}
{|J_{i,j}|}
},\vspace{1mm}\\
\omega\in J_{i,j}\quad \mbox{ for }\,
\tau(d_{i})=1,
\vspace{1.5mm}\\
\displaystyle{
\frac{-|J_i|\omega
+(d_{i,j-1}d_{j}
-d_{i,j}d_{j-1})}
{|J_{i,j}|}
},\vspace{1mm}\\
\omega\in J_{i,j}\quad \mbox{ for }\,
\tau(d_{i})=0.
\end{array}
\right.
\label{eqn:map}
\end{equation}
\begin{center}
\begin{figure}[t]
\begin{center}
\includegraphics[width=50mm]
{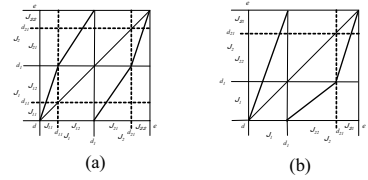}
\end{center}
\caption{An example of Kalman maps with (a) $4$ and (b) $3$ subintervals.}
\label{Kalman34}
\end{figure}
\end{center}
For simplicity, consider the case where $N_s=2$ and 
define the $N_s^2\times N_s^2$ matrix
\begin{equation}
{\widehat P}
=\left[
\begin{array}{cccc}
p_{11}&p_{12}&0&0\\
0&0&p_{21}&p_{22}\\
p_{11}&p_{12}&0&0\\
0&0&p_{21}&p_{22}
\end{array}
\right].
\label{eq:Markovtrhat}
\end{equation}
Let $\Lambda(\widehat{P})$ be the set of all eigenvalues of 
$\widehat{P}$. Then we get 
\begin{equation}
\Lambda(\widehat{P})=\Lambda(P)\cup0^{N_s^2-N_s},
\label{eq:Kalmanemmbed1}
\end{equation}
where $P$ is an $N_s\times N_s$ matrix with $N_s=2$, defined by
\begin{equation}
P=\left[
\begin{array}{cc}
p_{11}&p_{12}\\
p_{21}&p_{22}
\end{array}
\right].
\label{eq:Markovtr}
\end{equation}
Equation (\ref{eq:Kalmanemmbed1}) implies that a Markov chain is embedded
into the chaotic map $\tau(\omega)=\{\tau_{i,j}(\omega)\}_{i,j=1}^{N_s}$.
Figures \ref{Kalman34}(a) and (b) show an example of the Kalman map with 
 $4$ subintervals and a revised one with $3$ subintervals, respectively. 
\section*{Acknowledgment}
The authors would like to acknowledge the valuable and insightful comments and suggestions of the anonymous reviewers which improved the quality of the paper.
The authors would like to thank Prof. Yutaka Jitsumatsu for help in generating 
the simulations and figures.

\ifCLASSOPTIONcaptionsoff
  \newpage
\fi

\end{document}